\documentclass[12pt,preprint]{aastex}

\slugcomment{To appear in AJ}

\shorttitle{Pulsating Variables in Carina}
\shortauthors{Vivas \& Mateo}

\newcommand{\noprint}[1]{}
\newcommand{\figsetstart}{{\bf Fig. Set} }
\newcommand{\figsetend}{}
\newcommand{\figsetgrpstart}{}
\newcommand{\figsetgrpend}{}
\newcommand{\figsetnum}[1]{{\bf #1.}}
\newcommand{\figsettitle}[1]{ {\bf #1} }
\newcommand{\figsetgrpnum}[1]{\noprint{#1}}
\newcommand{\figsetgrptitle}[1]{\noprint{#1}}
\newcommand{\figsetplot}[1]{\noprint{#1}}
\newcommand{\figsetgrpnote}[1]{\noprint{#1}}

\begin{document}

\title{A Comprehensive, Wide-Field Study of Pulsating Stars in the Carina Dwarf Spheroidal Galaxy}

\author{A. Katherina Vivas} 
\affil{Centro de Investigaciones de Astronom\'{\i}a (CIDA)}
\affil{Apartado Postal 264, M\'erida 5101-A, Venezuela} 
\email{akvivas@cida.ve} 

\and

\author{Mario Mateo} 
\affil{Department of Astronomy, University of Michigan}
\affil{500 Church St. Ann Arbor, MI 48109, USA} 
\email{mmateo@umich.edu} 

\begin{abstract}

We report the detection of 388 pulsating variable stars (and some additional 
miscellaneous variables) in the Carina dSph galaxy over an area covering the full 
visible extent of the galaxy and extending a few times beyond its photometric 
(King) tidal radius along the direction of its major axis.   Included in this total are 
340 newly discovered dwarf Cepheids which are mostly located $\sim 2.5$ 
magnitudes below the horizontal branch
and have very short periods ($<0.1$ days)  typical of their class and consistent 
with their location on the upper part of the extended main sequence of the 
younger populations of the galaxy.
Several extra-tidal dwarf cepheids were found in our 
survey up to a distance of $\sim1\degr$ from the center of Carina. 
Our sample also includes RR Lyrae stars
and anomalous Cepheids some of which were found outside the galaxy's tidal radius as well.  
This supports past works that suggests Carina is undergoing
tidal disruption. We use the period-luminosity relationship for dwarf Cepheids to
estimate a distance modulus of  $\mu_0=20.17 \pm 0.10$ mags, in very 
good agreement with the estimate from RR Lyrae stars. 
We find some important differences
in the properties of the dwarf Cepheids of Carina and those in 
Fornax and the LMC,  the only extragalactic samples of dwarf Cepheids currently 
known. These differences may reflect a metallicity spread, depth along the line of 
sight and/or, different evolutionary paths of 
the dwarf Cepheid stars.

\end{abstract}


\section{Introduction}

The dwarf spheroidal (dSph) galaxies surrounding the Milky Way have proven to be complex objects
with extraordinary, and different, star formation histories \citep[e.g.][]{mateo98rev,grebel11}.
No two dwarf galaxies
in our neighbourhood are alike and the reason(s) why these galaxies present such a variety of star
formation and chemical enrichment histories is not well known today.
The understanding of the nature and evolution of these systems is fundamental for 
studying their role in the hierarchical process of formation of large galaxies like our own Milky Way. 

At $\sim100$ kpc of distance, Carina is one of the best known of  the "classical" 
dSph galaxies, in large part because of the rather extraordinary characteristics of its stellar 
populations. For example, the galaxy
presents a rich color-magnitude diagram (CMD) in which it is clear the existence of 
multiple stellar populations of ages around 11, 5 and 1 Gyr 
\citep{smecker96,hurley98,monelli03}. 
The CMD of Carina -- in particular its very narrow red giant branch -- suggests
that the spread in 
metallicity should be small in this galaxy \citep[e.g.][]{rizzi03,bono10},
though the well-known degeneracy between age and metallicity coupled with the internal age 
spread in Carina may be conspiring to mimic a small abundance spread.  
Spectroscopic measurements seem to show inconclusive and somewhat 
contradictory results on this issue \citep{koch06,helmi06}. 
In addition, Carina appears to be suffering significant tidal erosion from its interactions with the 
Milky Way as suggested from the existence of 
of Carina members beyond its photometric \citep{king62}
tidal radius \citep{majewski00,munoz06,battaglia12}.

Pulsating stars have been frequently used to study the stellar population in stellar systems.
The presence of RR Lyrae stars, for example, is an unequivocal signal that an old population 
($>10$ Gyrs) in the system exists. 
On the other hand, anomalous Cepheids, which are common in dSph 
galaxies but not in galactic globular clusters \citep{nemec94,clement01}, 
are usually interpreted as belonging to an 
intermediate/young stellar population 
\citep[][among others]{zinn76,bono97,dallora03,kuehn08,kinemuchi08,fiorentino12}.
Both types of pulsating stars have been found already in Carina 
\citep{saha86,dallora03}, consistent with what is know about the overall properties of its stellar 
populations.

Below the horizontal branch, the instability strip crosses 
an interesting location in the CMD of Carina where
pulsating stars from different evolutionary paths may coexist. There are not only
stars from the young and the intermediate age main sequence but also blue stragglers
from the old population of the galaxy. The nomenclature of the pulsating stars in this
part of the instability strip is confusing. We will follow \citet{mateo93} and will refer to them as dwarf 
Cepheids (DC). Some recent works have preferred to name them either 
as $\delta$ Scuti or as SX Phe depending 
on they belonging to a Population I or II, respectively. 
In complex systems such as Carina, 
the concepts of Pop I and II are not so easy to apply as it is not straightforward
to determine the metallicities of the population(s) in the region of the
instability strip.  Consequently, based on common definitions of the classes, a 
dwarf galaxy such as Carina could contain both $\delta$ Scuti and SX Phe stars 
though the data to distinguish them may not be readily available for the 
foreseeable future. For all these reasons, the collective name 'dwarf Cepheid' 
seems more appropriate when applied to variables in objects like Carina.

Properties of this kind of variable stars are discussed extensively in 
\citet{breger00}.
The key practical property of these stars is that DCs are short period variables 
($\sim0.03$ to 0.25 days). 
As a result, any observational study of these stars must maintain a short cadence 
while maintaining sufficient signal-to-noise to generate useful light curves for 
detailed discovery, classification and analysis.  Fortunately, SX Phe and a sub-group of the 
$\delta$ Scuti stars named HADS (high amplitude Delta Scuti)
have large amplitudes, amounting to several tenths of a magnitude, making these 
objects feasible targets with medium-size telescopes even at the distance of a galaxy such as Carina.

\citet{mateo98} presented the first detections of DCs in Carina. They studied three
small fields near the center of Carina and detected 20 of these 
short-period pulsators. Since DC stars
obey a period-luminosity relationship, \citet{mateo98} derived a distance to Carina
which was in very good agreement with estimates based on other types of variables 
and the properties of the galaxy's red giant branch. Given the pronounced 
mixture of stellar populations in Carina and, in particular, the importance of its 
intermediate age population, \citet{mateo98} speculated that the observed sample 
of DCs was likely just a small fraction of a likely large population of DCs in the 
galaxy.  In this work we extend the
previous observations by \citet{mateo98} and search
for DC in a much larger area, covering not only the whole galaxy but 
extending up to several tidal radii along the direction of the semi-major axis. 
Our aim is to use the DCs as possible tracers of extra-tidal features in Carina and, 
if found in sufficient quantities, as possible probes of the line-of-sight depth of 
the galaxy. RR Lyrae stars and anomalous Cepheids, which appeared naturally in 
our data, can also trace the extended structures of Carina.

We report here the discovery of 340 DC stars in Carina, which constitute one of the
largest extra-galactic samples of this type of stars known to date. 
Besides Carina, DCs have also been found in large numbers only in two other 
extragalactic systems, namely, Fornax \citep{poretti08} and the LMC 
\citep{garg10}. The different combinations
of age and metallicity of the stellar populations in these galaxies
allow us for the first time to compare the properties of DCs in distinct stellar 
systems. Such comparisons are useful
not only to shed light on the origin and specific frequency of DCs under 
different environments, but also, to study their use as standard candles and as
possible tracers of the 3D structure of galaxies.

\section{Observational strategy and data reduction}

Multi-epoch observations of eight fields around the Carina dSph galaxy 
were taken with the MOSAIC-II camera at the 4m Blanco Telescope at Cerro Tololo
Interamerican Observatory (CTIO), Chile, during an interval of 6 consecutive 
nights in December 2007.
Table~\ref{tab-fields} shows the central coordinates of the observed fields, the 
number of nights in which each field was observed and the total number of 
observations in the B and V bands. 
The complete survey covers the whole of the Carina dSph galaxy out to its classical 
\citep{king62} tidal radius of $28\farcm 8$ \citep{irwin95} and extends up to 
$\sim1\fdg 5$ from the galaxy  center in each direction along its semi-major axis. 
The outermost surveyed regions are located at about three tidal radii from the 
galaxy center.

\begin{figure*}[t]
\epsscale{0.9}
 \plotone{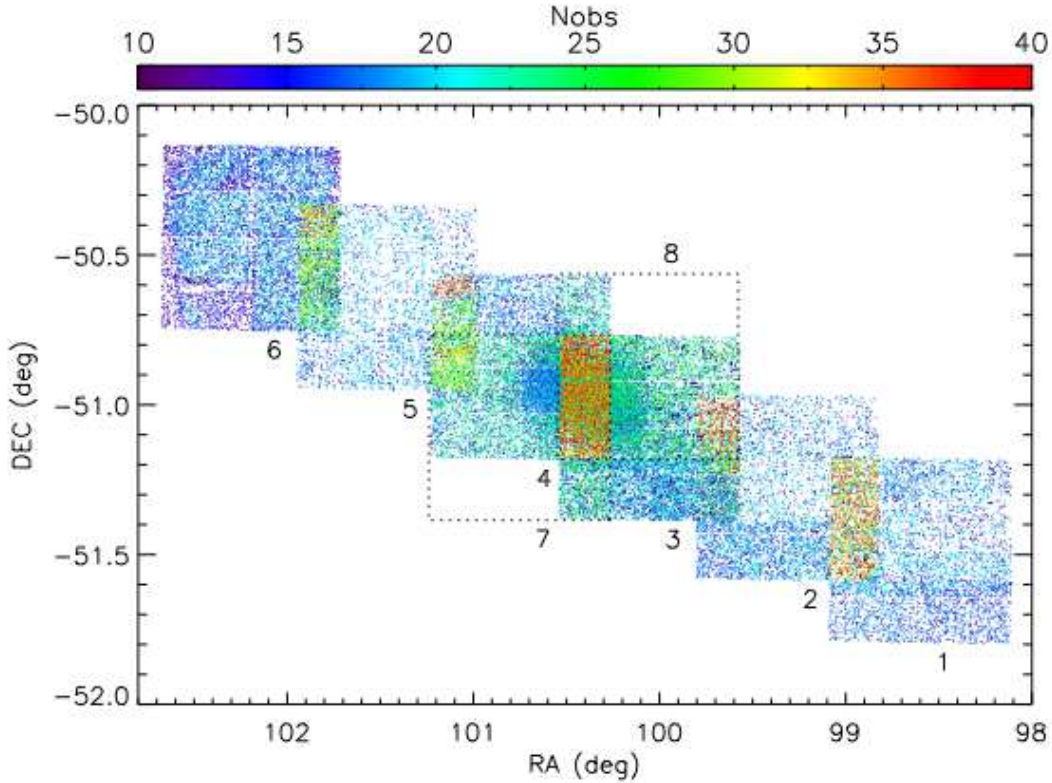}
 \caption{Total number of observations ($N_{obs}=N_V+N_B$) in all the survey region. 
 Numbers correspond to the Field number according to Table~\ref{tab-fields}.
 Fields 7 and 8, which had very few observations, are shown with dotted lines.}
 \label{fig-Nobs}
\end{figure*}

Since the main goal of these observations was to identify short period variables 
($<0.25$ d), the basic approach consisted in
obtaining repeated observations of the same field during a night. In general, 
continuous observations
of the same field for a period of $\sim$ 3-4 hours was obtained, typically 
alternating exposures in B and V. 
This strategy was chosen to span a full pulsation cycle of the longest-period DC 
variables anticipated to be present in Carina.  In practice, the multi-hour sequences 
covered one complete pulsation cycle for
stars with periods ($\lesssim 0.15$ d), and multiple (2-3) pulsation cycles for the 
majority of DC which are expected to have periods of the order of only 1 hour ($\sim 0.05$) d).
Additional observations of each field were taken as well in different nights to extend the temporal 
coverage over the full extent of the run.
Figure~\ref{fig-Nobs} shows the total number of observations ($N_B+N_V$) in the whole survey
region. Of these, $\sim60\%$ are V observations and the rest are B.
Fields had a sizeable 
overlap resulting in some regions having a larger number of observations.
Fields 7 and 8 were observed only twice in each band. Only the 
parts of those two fields overlapping other fields were used in this work.
There are at least 20 observations (total, in B and V) per star in all
the region although the central parts of the galaxy have significantly more 
observations, in some cases approaching 50 or more epochs. 
Depending on seeing conditions, exposure times varied from 480s to 600s in V 
and 500s to 600s in B, corresponding to about 10\% 
of the pulsation cycle of stars with periods of 0.05 d. 
This dataset is the same one used recently by \citet{battaglia12} to determine an age 
gradient in Carina. 

Individual frames were bias subtracted along rows using a constant derived from 
the image overscan region.   Flat-field images were obtained using dome and 
twilight observations (suitably dithered to avoid stellar contamination) and 
supplemented by dark-sky flats obtained from our observations of the off-center 
fields in Carina.  The latter were used only to apply very low spatial frequency 
corrections to post twilight/dome flattened images.  

\begin{figure*}
\epsscale{0.9}
 \plotone{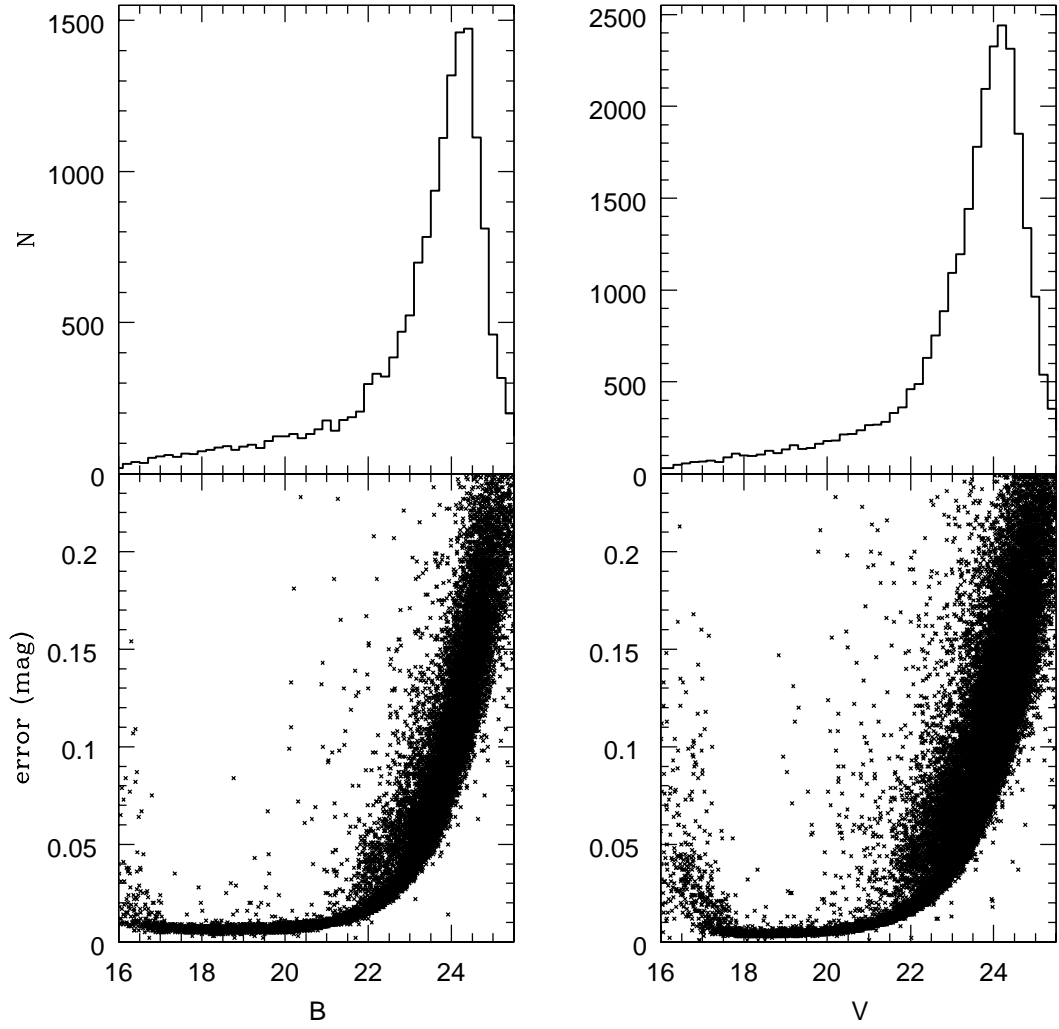}
 \caption{{\sl (Top)} Number of sources detected in individual B and V images of Field 1 as a 
 function magnitude. {\sl (Bottom)} Errors given by DoPhot for stars in the
 same images.}
 \label{fig-errorDAOPHOT}
\end{figure*}

The photometry was carried out using DoPhot \citep{schechter93}.  These 
reductions were carried out independently for each frame but using 
input coordinates  derived from photometry of the stacked images.   This resulted 
in more consistent photometric stability for non-variables and more uniform 
photometric catalogs for individual frames than reducing each frame 
independently.  
A typical distribution of photometric error as a function of V and B 
magnitudes is shown in Figure~\ref{fig-errorDAOPHOT}.  
We estimate that our data are typically complete to V, B $\sim 24.2$ since the number of
detected sources drops for fainter magnitudes (Fig~\ref{fig-errorDAOPHOT}).
Saturations signs start to be seen for magnitudes brighter than B, V $\sim 17$.

From the photometry of the individual frames we were able to to obtain time series 
for all stars in the region.

\section{Selection of variable stars}

The MOSAIC-II camera consists of eight $2048 \times 4096$ CCDs in a $4 \times 2$ array
to cover a nearly square $36$ arcmin$^2$ field. 
In order to ensure the best relative photometry in the time 
series for each star we treated each CCD individually. For each CCD, in each field, and each band, 
we selected a
reference image, usually the one with the best seeing and hence the largest number of detected stars. 
Based on their celestial coordinates,
we then matched stars
in the reference image with all other observations not only from the same field but also from
overlapping regions in other fields. A zero point difference was estimated after applying an iterative
$3\sigma$ clipping using all stars with photometric errors $<0.05$ mags. 
The $3\sigma$ clipping avoided variable objects or spurious measurements to enter in the
calculation of the zero point, which was then applied to all stars in each image. Even in 
images with only partial overlap with the reference image, the number of stars entering in the
calculation of the zero point was large enough (usually over 100) to make this value
very robust. This normalization procedure resulted in reliable time series for all stars in the 
fields from which we could then investigate variability. 

\begin{figure*}
\epsscale{0.9}
 \plotone{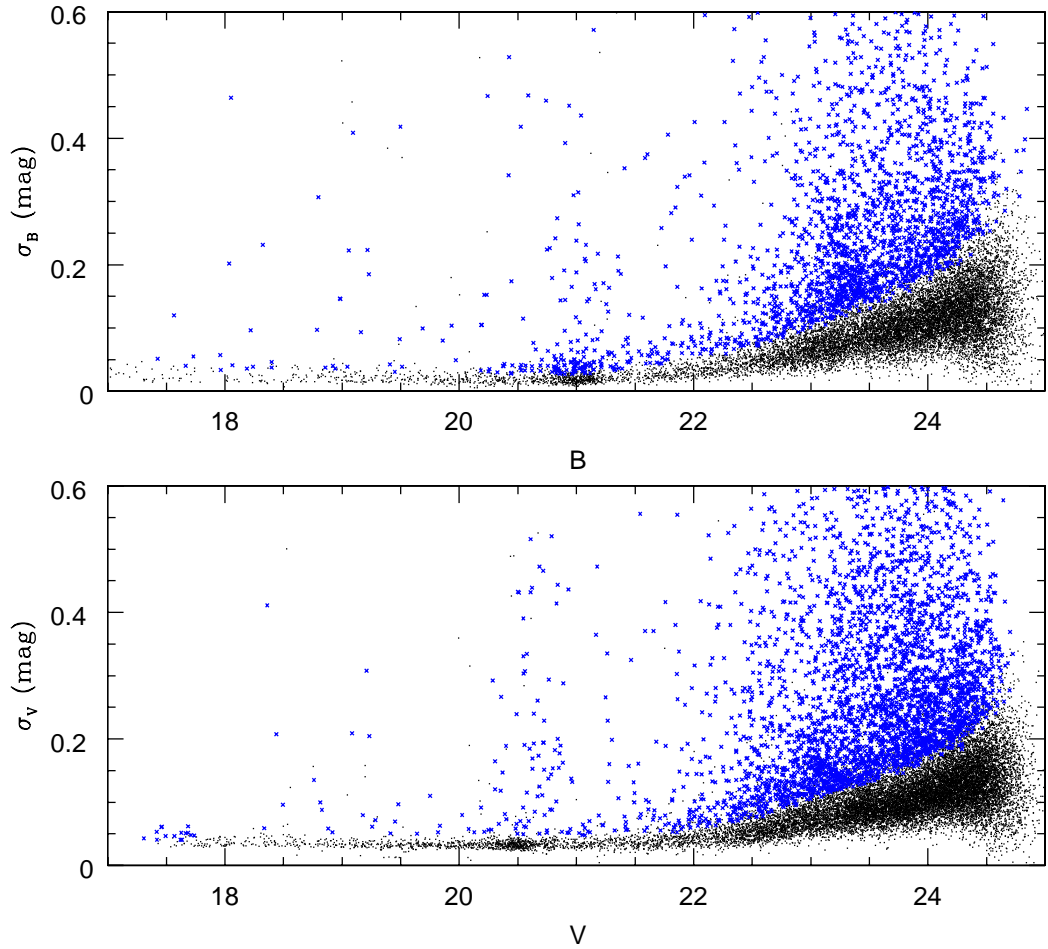}
 \caption{Standard deviation of the magnitudes of all stars in Field 3, CCD7, in B ({\sl top}) and
 V ({\sl bottom}) bands. This particular frame contains stars in the innermost part of the Carina 
 dSph. Each star was observed up to 22 times in B and 30 times in V. Crosses in both 
 diagrams indicate stars selected as variable by the $\chi^2$ test.}
 \label{fig-errors}
\end{figure*}

Figure~\ref{fig-errors} shows the resulting standard deviation of the B and V magnitudes of stars
for one of the fields
(Field 3, CCD 7) which covers the innermost, hence most crowded, parts of Carina. The 
photometric error at the 
bright end amounts to $\sim0.02$ and $0.03$ mags in the B and V band, respectively. 
These numbers go
down to $\sim0.01$ mag in the outermost fields where crowding is far less of an issue. 
For the stars of interest, the faint limit is similar in both photometric bands.
Stars have in average a standard deviation of 0.1 mag at magnitude 23.8  in both bands. 
The main locus of stars shown in Figure ~\ref{fig-errors} (and similar ones for others
fields/CCDs) empirically define the photometric errors as a function of magnitude and may be 
compared to the photometric errors estimated by DoPhot in Figure~\ref{fig-errorDAOPHOT}.
This run of photometric uncertainties for non-variable stars will be assumed for all 
stars from all fields analysed in this study. 

To identify variable objects we performed $\chi^2$ tests on each star 
independently for both the B and 
V bands. With this test we could flag as variable stars those stars whose variations in brightness are 
not likely due to photometric errors alone, by requesting that the probability distribution 
function be a small number, $P(\chi^2)<0.01$ \citep[see for example][]{vivas04,watkins09}. 
The flagged stars are shown in the example field of Figure~\ref{fig-errors} as crosses.
We required that stars should be flagged as variable in both B and V bands, a condition that 
eliminated many spurious or otherwise marginal cases.

\begin{figure*}
\epsscale{0.9}
 \plotone{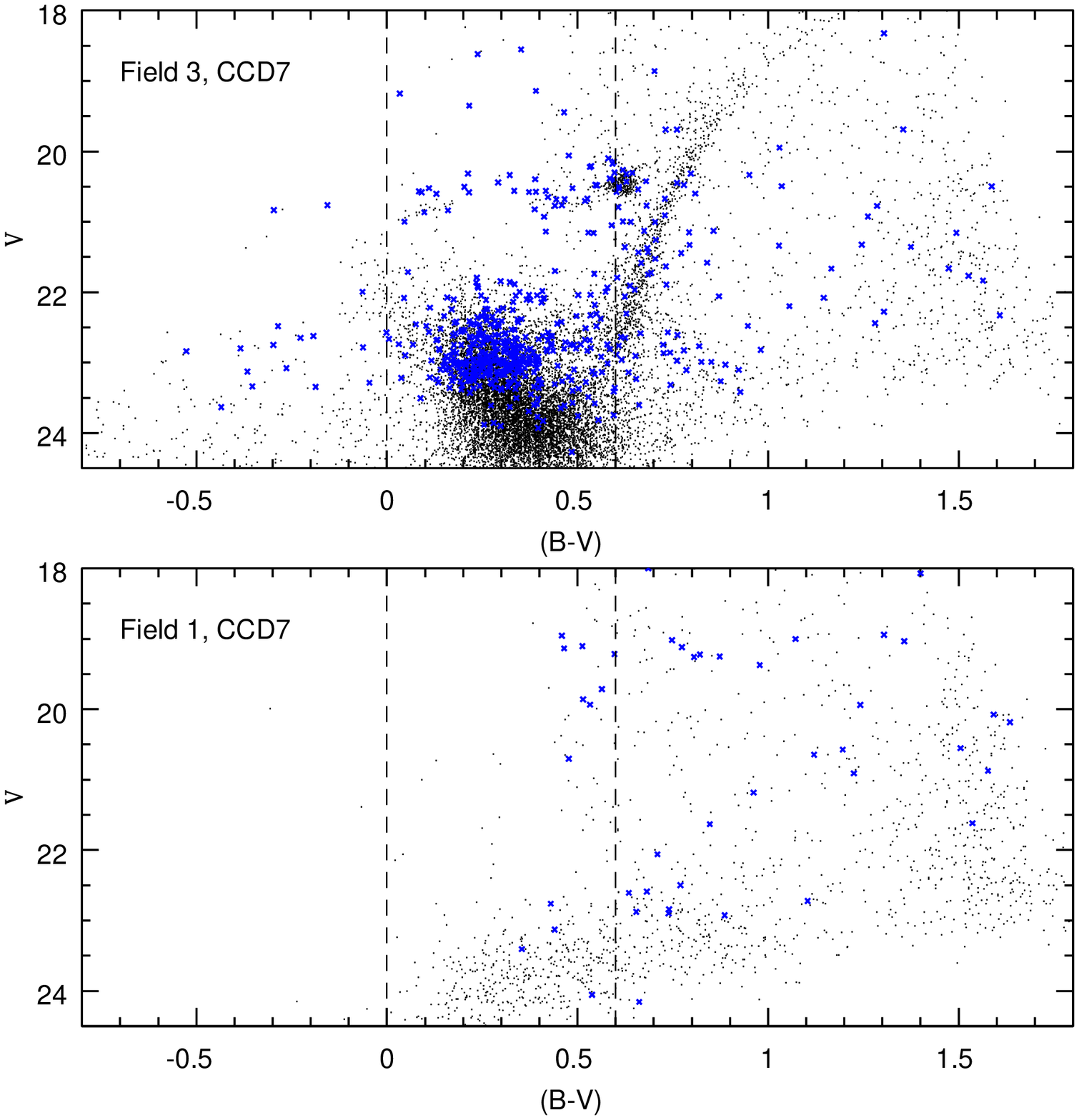}
 \caption{CMD in one of the fields near the center of Carina ({\sl top}) and one outer 
 field ({\sl bottom}). Variable stars are shown as crosses. Variable stars within the dotted 
 vertical lines were checked for periodicity.}
 \label{fig-cmd_var}
\end{figure*}

Figure~\ref{fig-cmd_var} shows the CMD in  one of the central fields in Carina (Field 3, CCD 7) and one
of the outermost ones (Field 1, CCD 7). Variable stars identified with the method just described
are shown with $\times$ symbols. 
The upper panel clearly shows the very well known features of Carina: a double 
turnoff, a clear horizontal branch and a prominent red clump. 
It is also obviously apparent from this diagrams that Carina has a large population of variables.
In particular, it is  remarkable the large concentration of variable stars located in the upper 
main sequence, all of which are clear candidate DC variables.
There are also variable objects at the horizontal branch, which are 
expected to be RR Lyrae stars. 
Brighter blue variables may be anomalous Cepheids 
known to exist in the Carina stellar population \citep{saha86,nemec94,dallora03}. 
In the particular field shown in the upper panel of Figure~\ref{fig-cmd_var} , 602 ($\sim4\%$) of the
stars were flagged as variable using our $\chi^2$ criteria described above. 
The field in the lower panel is located at $\sim2\degr$ in RA
from the center of Carina (Field 1, CCD7). 
The features of the galaxy are not easily discernible here due to the low 
number of stars (only the area covered by one CCD is shown). 
Previous studies have shown, however, that the galaxy appears to extend this far 
\citep[][among others]{majewski00,munoz06,battaglia12}. The number of variables is lower in this
field but nonetheless there are several ones located within the color-magnitude range more 
clearly defined by the variables at the center of the Carina.

Pulsating stars are expected to be located exclusively within their respective 
instability strip.   To pare down the sample to include the most viable pulsating  variable 
candidates, we have restricted our search for periodic variability to those  stars with mean 
photometric properties that place them within a generous color range enclosed by the
dotted lines in Figure~\ref{fig-cmd_var}
($0<(B-V)<0.6$). These limits in 
color are loose enough to give room for photometric errors and possible variable extinction.

\section{Periodic variables}

The next step for the correct identification of the variable stars is the determination of periodicity.
Our dataset does not have a large number of epochs which may be a problem to determine 
periods, though this is somewhat mitigated by the multi-hour (and multi-period, for DCs) 
stretches of observations in our dataset for each field.  We used the
combined B and V observations in conjunction
with the \citet{lafler65} algorithm. For each star we calculated the string length for a given trial period,
defined by \citet{lafler65} as:

\begin{equation}
\Theta(\lambda) = \frac {\sum_i {(m_i-m_{i+1})^2}} {\sum_i{(m_i-\bar{m})^2}}
\end{equation}

\noindent
where the magnitudes $m_i$ have been sorted by increasing phase for that particular trial period.
If a star is periodic, the string length in both B and V bands should be
a minimum when the data is phased with the correct period. For each trial period we 
calculated the string length in each band and then combined them
weighting by the number of observations in each band \citep{stetson96,watkins09,mateu12}:

\begin{equation}
\Theta = {N_V \Theta(V) + N_B \Theta(B) \over {N_V+N_B}}
\end{equation} 

We imposed a requirement of a minimum of 5 observations in each band, and at least 15 observations in 
total ($N_V+N_B$) to use this approach to test a star for periodic variablity.
To be able to study possible aliasing, for each star we selected the first three minima of $\Theta$ in the
trial range. We checked visually the phased light curves of a star with these three periods if
the parameter $\Lambda>2.0$. This parameter is defined in \citet{lafler65} as the ratio between
the string length away from and at the right period. The larger this number, the deeper the minimum
and the most likely for the period to be real. The value of 2.0 was chosen arbitrarily after
examining the results for a few fields so that we included a manageable number of false positives
in our sample while minimizing the number of missed true periodic variables.  
The constraint is loose enough to allow the recognition of
periodic variables even in stars for which a small number of epochs are available.
If the trial period is the right one, the light curve should look smooth and coherent in both B and V bands.
Since our main goal
was to explore the faint pulsating variables we first searched for periodicity in the range 0.03 to 0.2 d.

The bright variables in Carina have been explored extensively by \citet{saha86} and
\citet[][hereafter D03]{dallora03}, although the latter in a 
region smaller than the one studied in this work. Our time sampling is not ideal for 
exploring periods $>0.5$d like the ones for RR Lyrae stars or
anomalous Cepheids. Nonetheless, we tried the period search also in the range 0.2 to 0.9 d, knowing
that the completeness, due to our time sampling, may be low for stars in this range of periods.

\begin{figure*}[t]
\epsscale{0.7}
 \plotone{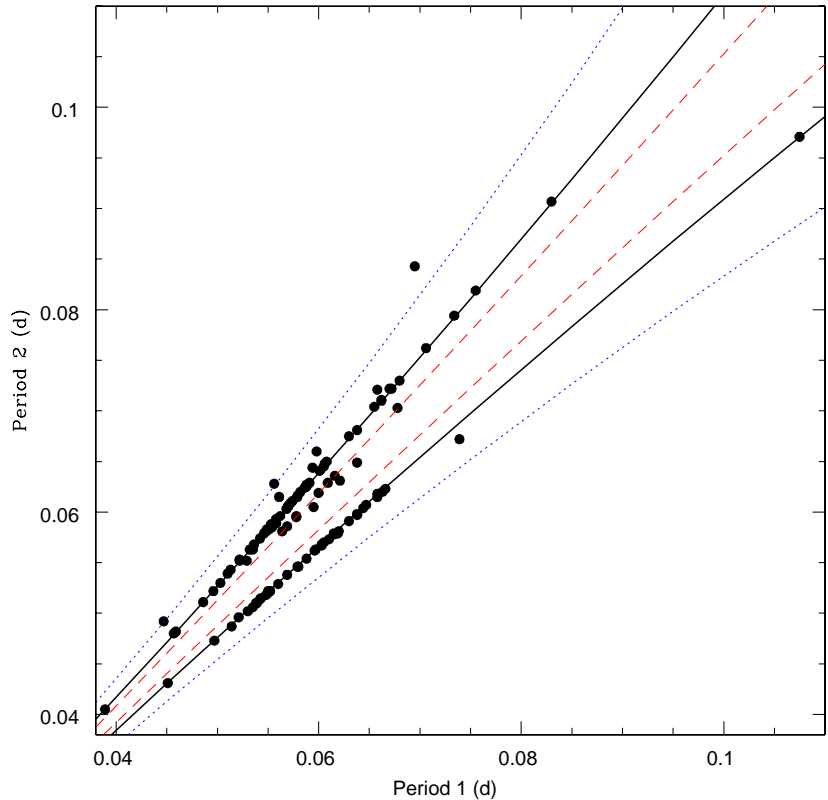}
 \caption{Relationship between the two possible periods found for 116 short-period variables. The lines
 show the expected relation given by equation~\ref{eq-alias} for $p=1$ (solid lines),
 $p=2$ (dashed lines) and $p=1/2$ (dotted lines).}
 \label{fig-alias}
\end{figure*}

Visual examination of the resulting light curves revealed aliasing affects in many of the short 
period variables. 
Indeed, for $\sim1/3$ of the short period variables ($<0.02$ d),
we found reasonable light curves with two different periods. According to \citet{lafler65},
the most common alias periods ($\Pi$) are found when $p={1,1/2,2}$ in the equation:

\begin{equation}
\Pi^{-1} = P^{-1} \pm {1 \over p}
\label{eq-alias}
\end{equation}

\noindent
where $P$ is the true period of the star. 
Figure~\ref{fig-alias} shows the relation between the two periods
found for some stars in our sample.  
We also plotted the lines of the $\pm 1$-day alias ($p=1$, solid line), 
the 2-day, and $1/2$-day aliases ($p=2$ and $1/2$, dashed and dotted lines respectively) 
given by equation~\ref{eq-alias}. 
It is clear that the majority of the stars lie along the solid
lines, confirming that the most common alias in our data
is, as expected, the $\pm 1$-day alias. The other two aliases are nonetheless also present.  
However, at 
the short period range of these stars, aliasing it is not a significant problem for our analysis 
since the differences 
between the real periods and their aliases
are generally quite small, typically $<6\%$.
For the remainder of this paper we assumed the true period to be
the one giving the best light curve to the eye, which is usually the one with the largest value
of $\Lambda$. Nonetheless, we report all suspected aliases as well.

\begin{figure*}
\epsscale{0.84}
 \plotone{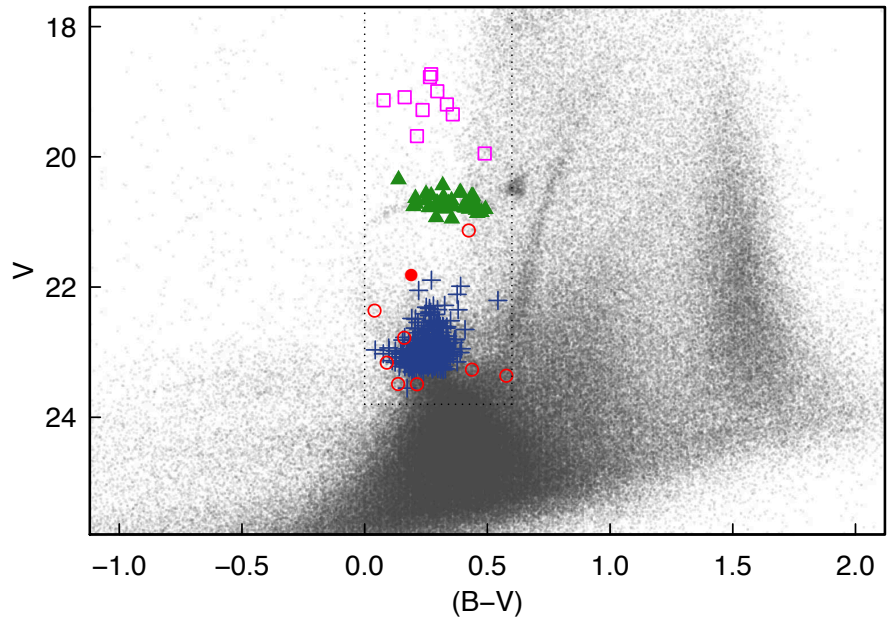}
 \caption{CMD of all stars in the survey. The colored symbols indicate the periodic variables 
 found in this study: dwarf cepheids ($+$ symbols),
 RR Lyrae stars (solid triangles), anomalous cepheids (open squares)
 and miscellaneous stars (circles). Among the latter, the solid circle indicates the location of
 a clear case of an eclipsing binary. The dotted lines enclose the region of the CMD in which
  we search for periodicity among the variables.}
 \label{fig-cmd}
\end{figure*}

\begin{figure*}
 \plotone{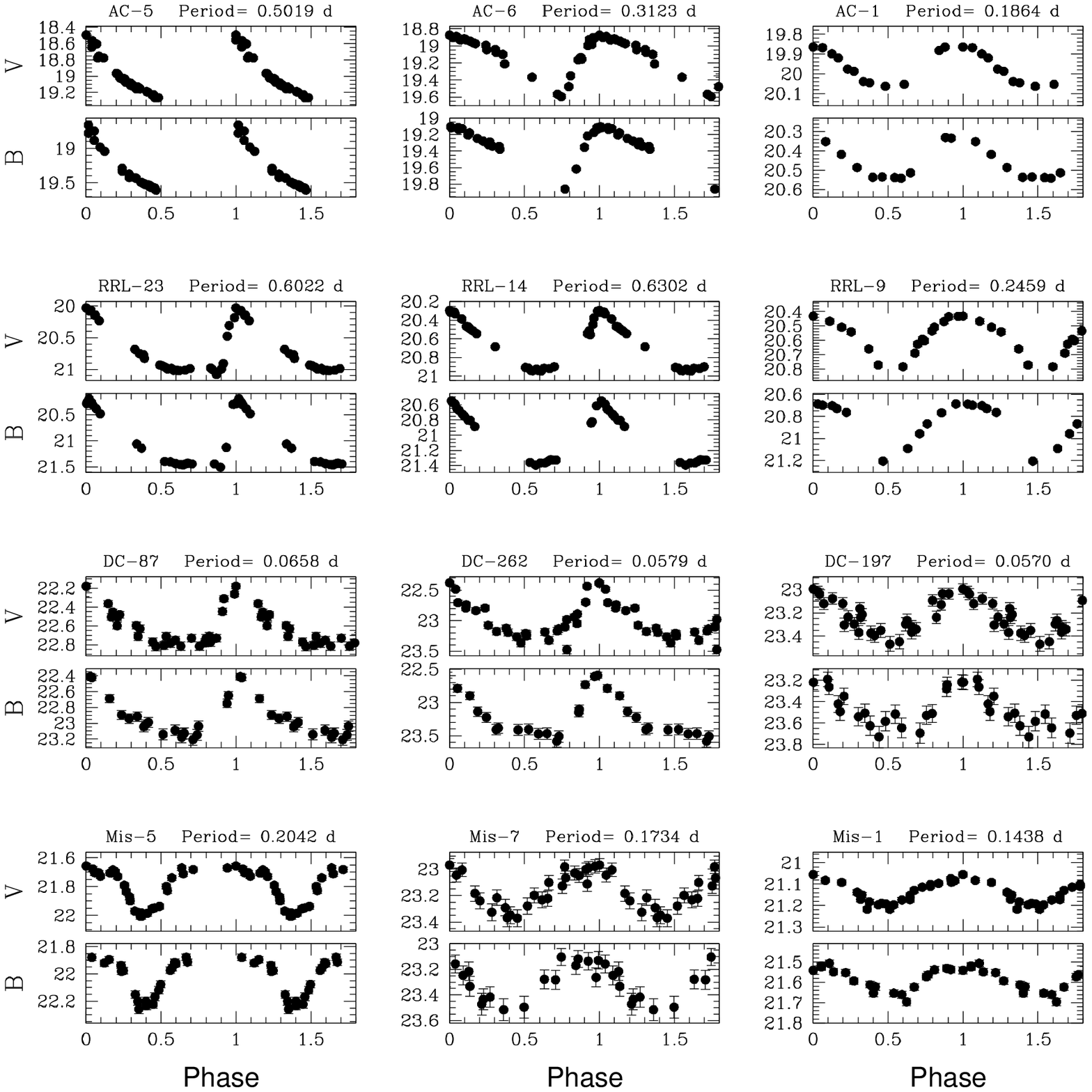}
 \caption{Examples of light curves. From top to bottom, each row presents examples of anomalous Cepheids,
 RR Lyrae stars, DC and stars in the miscellaneous group.}
 \label{fig-lc}
\end{figure*}

We found a total 397 periodic variable stars, in the range from 0.03 to 0.9 d, in Carina. 
Based on their location on the CMD (Figure~\ref{fig-cmd}), 
and their periods and light curve shapes,
we separated the periodic variables into four different groups 
which are seen in Figure~\ref{fig-cmd} with
different symbols.
The majority of the periodic 
variable (340) are strongly concentrated in the upper part of the main sequence ($+$ symbols). 
We take these stars to be DCs which may encompass both SX Phe and large-amplitude 
$\delta$ Scuti stars. 
There are also many number of stars (38) at the level of the horizontal branch which 
we confidently presume to be RR Lyrae 
stars. At even brighter magnitudes we found 10 periodic variables which we identify as 
anomalous Cepheids, which have been known to exist in Carina \citep[][D03]{saha86}. 
There is finally a group of 9 stars whose 
identification is more uncertain an we classify as {\sl miscellaneous}. 
Each one of these 
groups are discussed with more detail in the following sub-sections. 
Examples of light curves are shown in Figure~\ref{fig-lc}. 
Because of the large number of variables,
the full set of light curves is available as online-only material (Appendix~\ref{ap-lc}). 

\begin{figure*}
\epsscale{0.9}
 \plotone{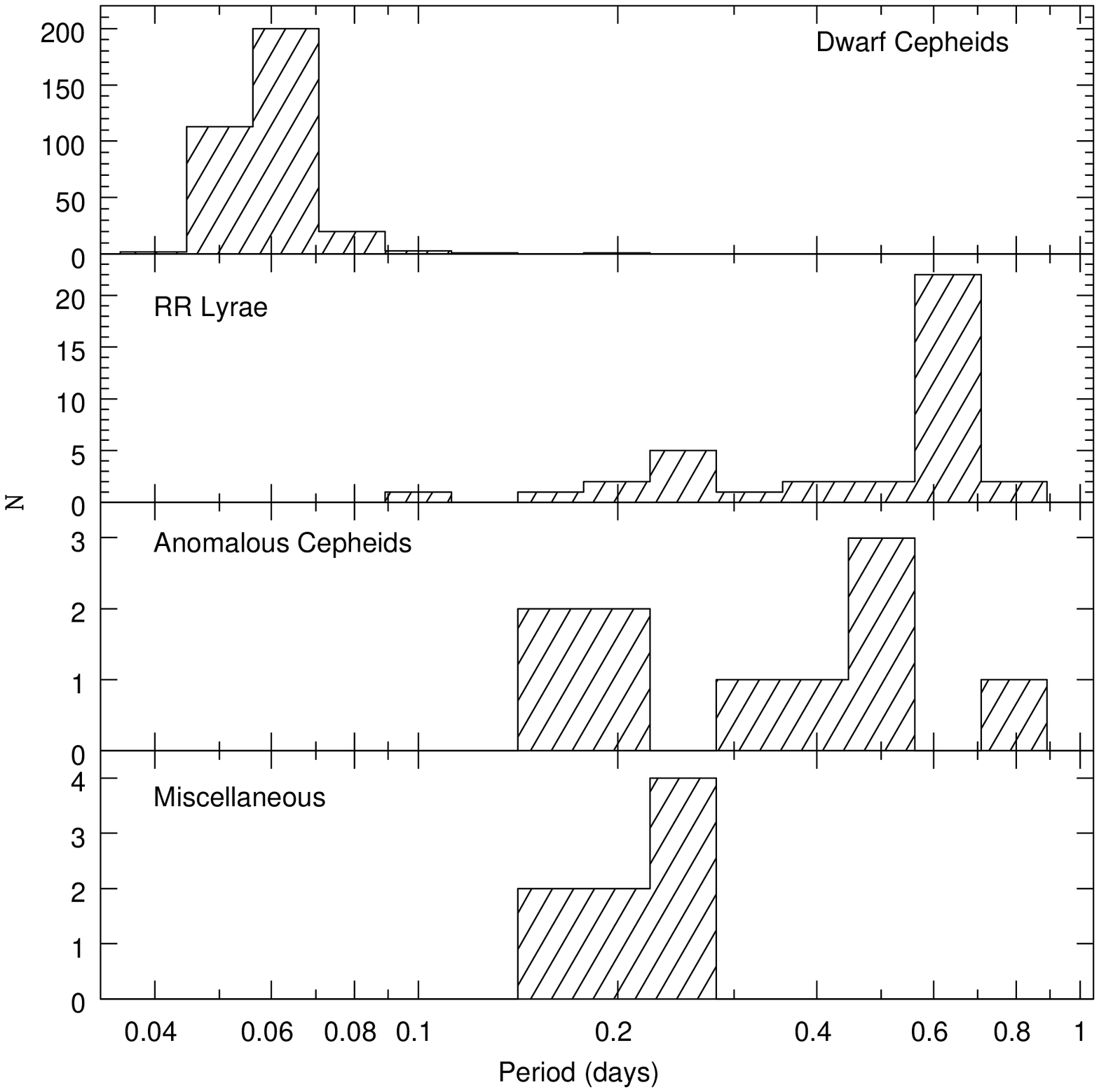}
 \caption{Period distribution of stars in the four groups of variables identified in this work.}
 \label{fig-periods}
\end{figure*}

The period distributions of all variable stars are shown 
in Figure~\ref{fig-periods}. 
DCs, by far the most numerous of the sample, were found to have periods
between 0.03 and 0.18 d, strongly peaked at a period of 0.06 d.
Both RR Lyrae and anomalous Cepheid stars span a period range between 0.1 and 0.9 d.
Finally, the stars in the miscellaneous group, which are dispersed in the CMD 
but below the horizontal branch, 
have (uncertain) periods longer than the DC stars, ranging from 0.14 to 0.24 d. 

\subsection{RR Lyrae Stars}

\begin{figure*}
\epsscale{0.8}
 \plotone{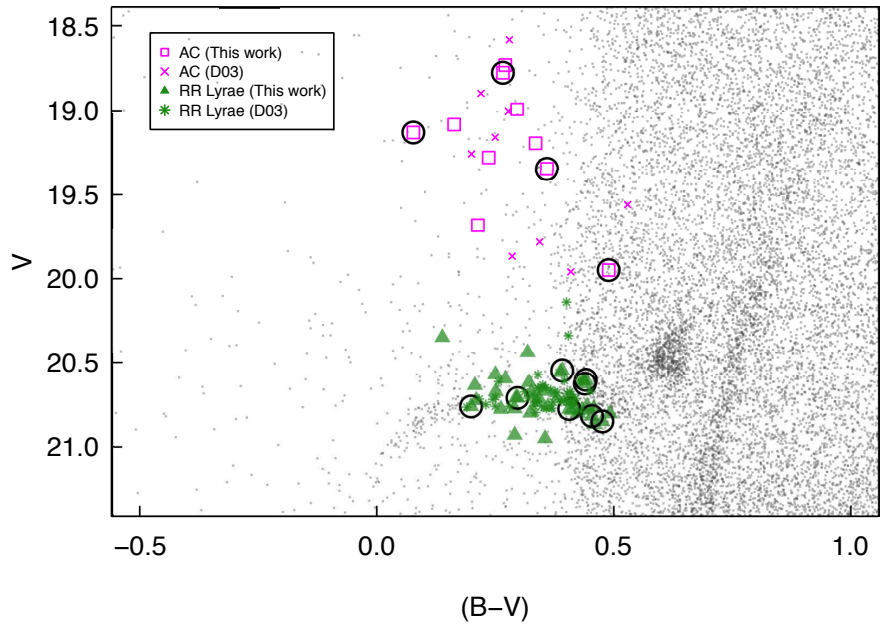}
 \caption{Enlargement of the CMD in the region o the bright variables. 
 RR Lyrae stars and anomalous Cepheids
 identified in this work are indicated by 
 the triangles and open squares respectively. New discoveries are enclosed by
 open circles. Stars in D03 not recovered in this work are shown with asterisks 
 ans $\times$ symbols for RR Lyrae stars and anomalous Cepheids respectively.}
 \label{fig-CMDb}
\end{figure*}

We detected 38 RR Lyrae stars in our fields, 27 of them were classified as type {\it ab} (RRab) and 11 as type {\it c} (RRc).
Their properties are listed in Table~\ref{tab-RR}. Examples of the light curves of 
2 RRab (left and middle
column) and 1 RRc (right column) are shown in the second row of Figure~\ref{fig-lc}. 
As mentioned before, D03 searched exhaustively for bright variables in Carina. They found 75 
RR Lyrae stars. Even with our non-ideal sampling for this type of stars, 
we recovered 30 of them. The ID number
in D03, together with their type classification and period, is included in Table~\ref{tab-RR}. 
The 20 type RRab in common with the sample in D03 agree in their periods within 0.01 d. 
We also recovered the correct dominant period for 1 type d star. However, we
did not attempt to find double
mode pulsators, hence, type d stars in 
D03 were classified here as either type ab or c.
We recovered aliases for all the type c, and the two remaining type d stars in common. 
Type c stars
are most susceptible to aliasing problems since the $\pm 1$ d alias correspond to periods within
the expected range for these type of stars \citep[e.g.][]{vivas04}. 
Three stars (RRL-6, RRL-10 and RRL-29), which were
classified by us as RRc, were labelled as RRab in D03. Most likely the disagreement is due to our poor phase 
coverage for some of these star. Since D03 had lightcurves with a significantly higher number of epochs,
we recommend the use of their periods and classification for the stars in common.
Figure~\ref{fig-CMDb} shows a zoomed version of the CMD in the region of the bright variables.
All of the RR Lyrae stars concentrate in a narrow range of magnitudes. The mean magnitude of the
38 stars detected in this work is $V=20.71$, with a standard deviation of only $0.12$ mags. This
value agrees well with the mean found by D03 for their whole RR Lyrae sample, 
$V=20.68 \pm 0.06$.  

\begin{figure*}
\epsscale{0.76}
 \plotone{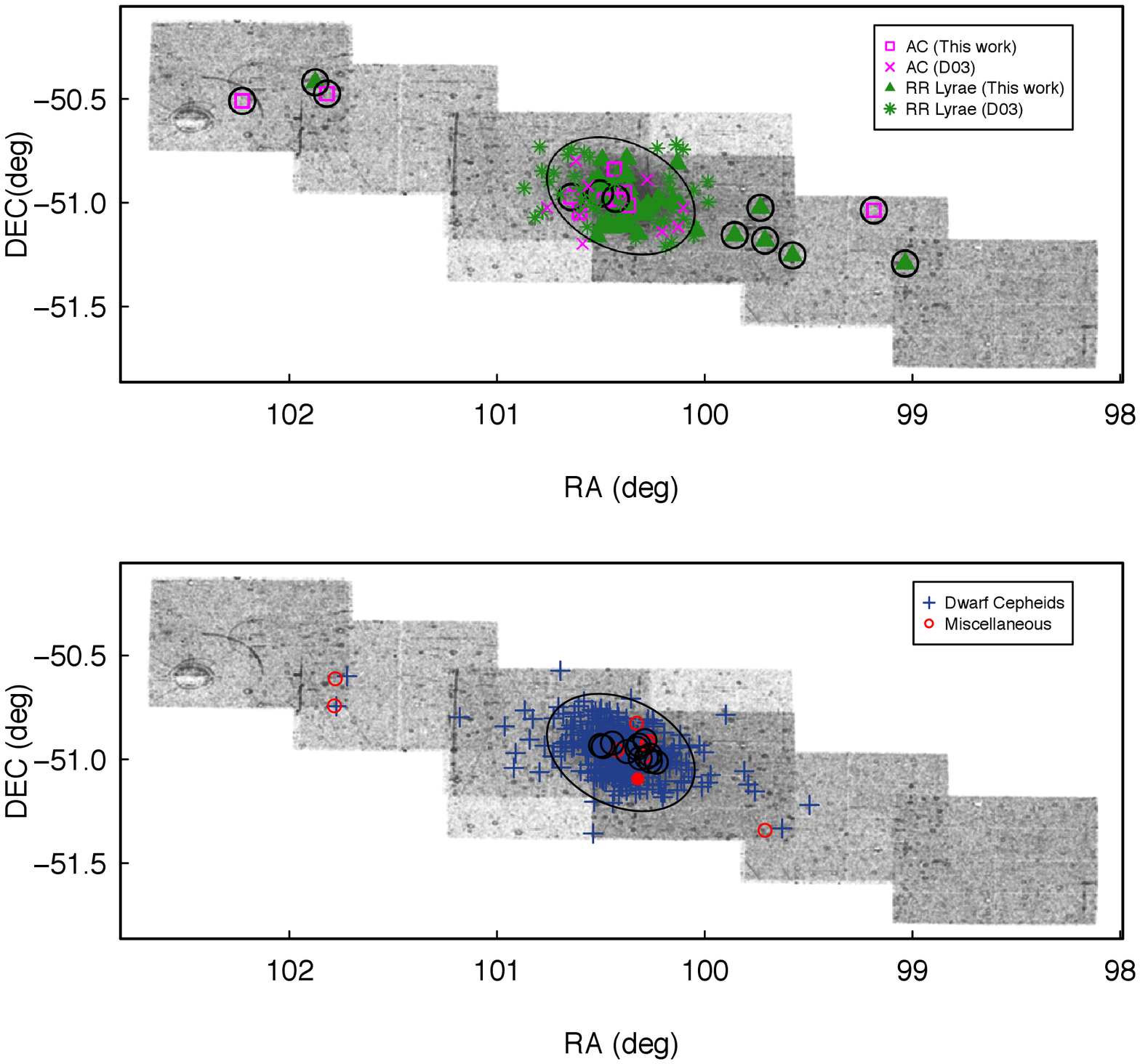}
 \caption{Distribution in the sky of the periodic variables detected in this work. The top panel
 shows the bright variables (RR Lyrae stars and anomalous Cepheids) while the bottom panel shows
 the faint ones (DCs and miscellanous stars).
 Symbols are the same as in Figure~\ref{fig-CMDb} and ~\ref{fig-CMDf}. 
While open black circles in the top diagram enclose the new RR Lyrae stars and anomalous Cepheids discovered in this work, in the lower diagram the same symbol is used to mark the
previously known DC from \citet{mateo98}.  
 In both plots the ellipse 
 represents the tidal radius ($22\farcm 54$), ellipticity ($0.32$) and position angle ($64\degr$) of
 Carina, according to \citet{walcher03}. }
 \label{fig-sky}
\end{figure*}

By looking at the spatial distribution of the RR Lyrae stars in our sample 
(Figure~\ref{fig-sky}), it is clear
that the detection of this type of stars was only possible where we had overlap 
among fields (see Figure~\ref{fig-Nobs}), and hence we had a larger number of epochs.
Stars in the D03 catalog that we did not recover were located mostly
outside of the overlap regions and hence, they did not have enough epochs and phase coverage 
in our survey to be recognized as RR Lyrae stars. We recall that in the regions with no overlap observations
were limited to several exposures within a period of $\sim 4$ hours, with only 1 or 2 additional
exposure on a different night. In most circumstances, this will not sample a significant part of
the light curve of
an RR Lyrae star, most especially, for the ones with longer period.

Eight new RR Lyrae stars in Carina are reported in this work; they are marked with 
open circles
in Figures~\ref{fig-CMDb} and \ref{fig-sky}.  Six of them are located outside the field
studied by D03, and far from the center of the galaxy.  

\subsection{Anomalous Cepheids}

We found 10 periodic variables brighter than the horizontal branch (see Figures~\ref{fig-cmd}
and~\ref{fig-CMDb}) which we identify as anomalous Cepheids in Carina. 
Six of these had been previously classified as such
by D03.   
Out of these 6 stars, we recovered the period within $<4\%$ for three of them
(AC-2, AC-3 and AC-5); 
for two of them we recovered alias periods (AC-6 and AC-7); 
finally, for one star (AC-4) 
the period reported by
D03, 1.16 d, was outside the range we studied in this work. However, we 
found a period for that star of 0.55 d, which produces a nice and smooth light curve. We
notice that one period is close to the harmonic of the other. 
As with the RR Lyrae stars, since the light curves of D03 contains
more epochs than ours, their periods are likely the correct ones for these longer
period variables.
Examples of three light curves of anomalous Cepheids are shown in the top row of
Figure~\ref{fig-lc}.

We report finding four new anomalous Cepheids, three of them located outside the region 
studied by D03 and some of them up to $\sim1\degr$ from the center of Carina (Figure~\ref{fig-sky}). 
Table~\ref{tab-AC} contains the data for all the anomalous Cepheids found in this work. The table includes
the ID and period determined by D03.
With these 4 new stars the total number of known anomalous Cepheids in Carina is 10.

\subsection{Dwarf Cepheids}

This is by far the most numerous kind of variables we found in our study. 
This partly reflects our observing strategy which was designed specifically to 
maximize the detection and recovery of these types of short period variables,
but also the sheer high frequency of DCs in Carina. 
With the identification of 340 DC stars, Carina is one of the stellar system with the largest known number 
of this type of stars, only surpassed in total number by the LMC \citep{garg10}. 
The only other external system to the Milky Way known to contain a large number of 
DCs is the Fornax dSph galaxy \citep[85 DCs,][]{poretti08}.

\citet{mateo98} looked for DCs in three small fields in Carina 
($3\farcm 9 \times 3\farcm 9$ each), finding 20 of these variables. We confirmed
12 of them in this work for which we recovered periods within $10^{-3}$ d (this difference includes
that we may have recovered the 1-day alias for some stars). The stars in common with the
sample in \citet{mateo98} 
are V4, V5, V9, V10, V12, V13, V14, V15, V16, V17, V18 and V20 (these IDs correspond to the
ones assigned by Mateo et al). Out of the 8 stars which were not
recognized as dwarf cepheids, 2 of them (V6 and V8) show variability in our lightcurves but no 
period within the searched range was found. 
The variability of the remaining 6 stars
is more uncertain in our data. Some stars (V3, V7 and V19) present either variability at the
level $P(\chi^2)<0.1$ (instead of $P(\chi^2)<0.01$), sometimes in the V band only, or no 
variability at all (in the case of V1, V2 and V11).

\begin{figure*}
\epsscale{0.85}
 \plotone{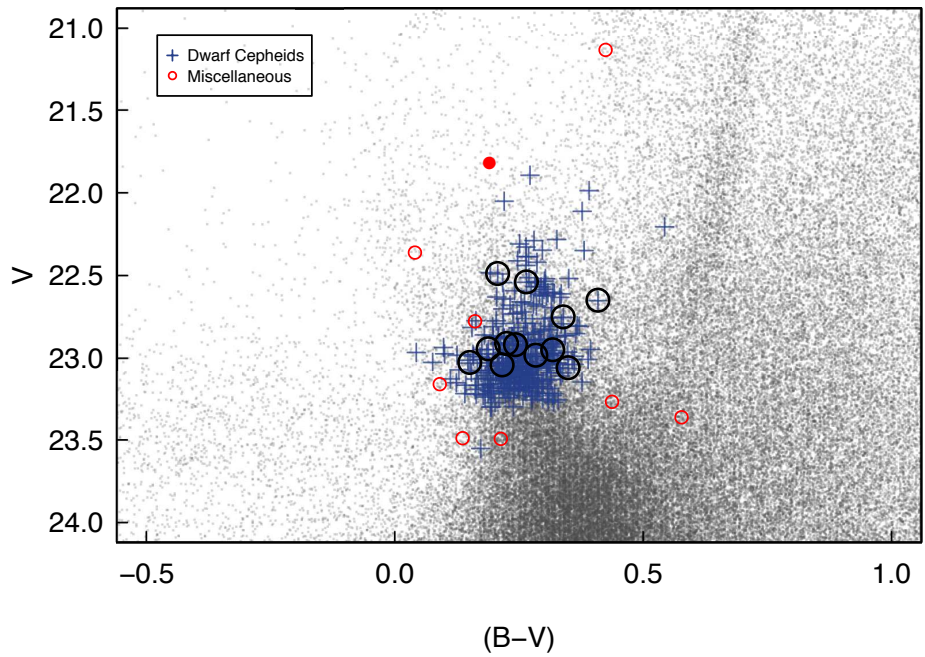}
 \caption{Enlargement of the CMD in the region o the faint variables. DC stars are indicated by 
 the $+$ symbols. 
 The clear example of an eclipsing binary is the solid circle, while the rest of the 
 stars in the miscellaneous group are shown with small open circles.
 Previously known DC stars in Carina \citep{mateo98} are enclosed by thick open circles.}
 \label{fig-CMDf}
\end{figure*}

Figure~\ref{fig-CMDf} shows a zoom of the CMD in the region of the faint variables. The location of the
previously known DC stars from \citet{mateo98} are indicated with thick open circles. 
The 340 DC stars are
strongly concentrated in color, exhibiting a mean sample
$(\langle B \rangle - \langle V \rangle)$ of $0.26\pm 0.06$ mag ($1\sigma$).
On the other hand, there is a relatively larger spread in brightness. Although 
most of the stars clump around magnitude
$V\sim 23.1$, there is a large tail toward brighter magnitudes, with a secondary peak at 
$V=22.6$, and maybe a third concentration at $V=22.35$. We will see below that these different 
peaks may be associated with different pulsation modes of the DC stars.

\begin{figure*}
\epsscale{0.9}
 \plotone{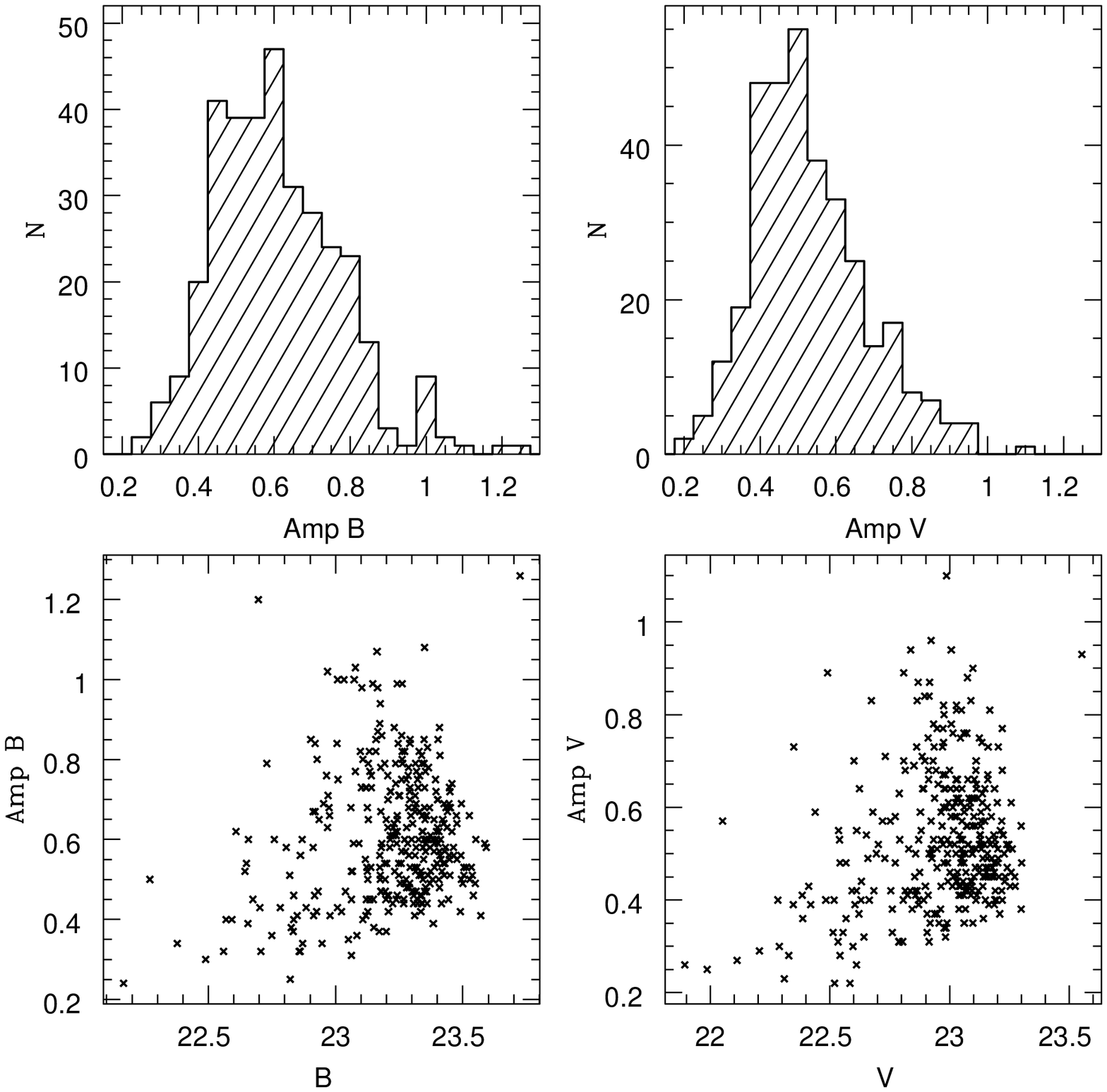}
 \caption{{\sl (Top):} Distribution of the amplitudes of the DC stars in the B and V band. {\sl (Bottom):}
  Amplitudes in B and V as a function of the mean magnitude. }
 \label{fig-DCamp}
\end{figure*}

\begin{figure*}[t]
\epsscale{0.9}
 \plotone{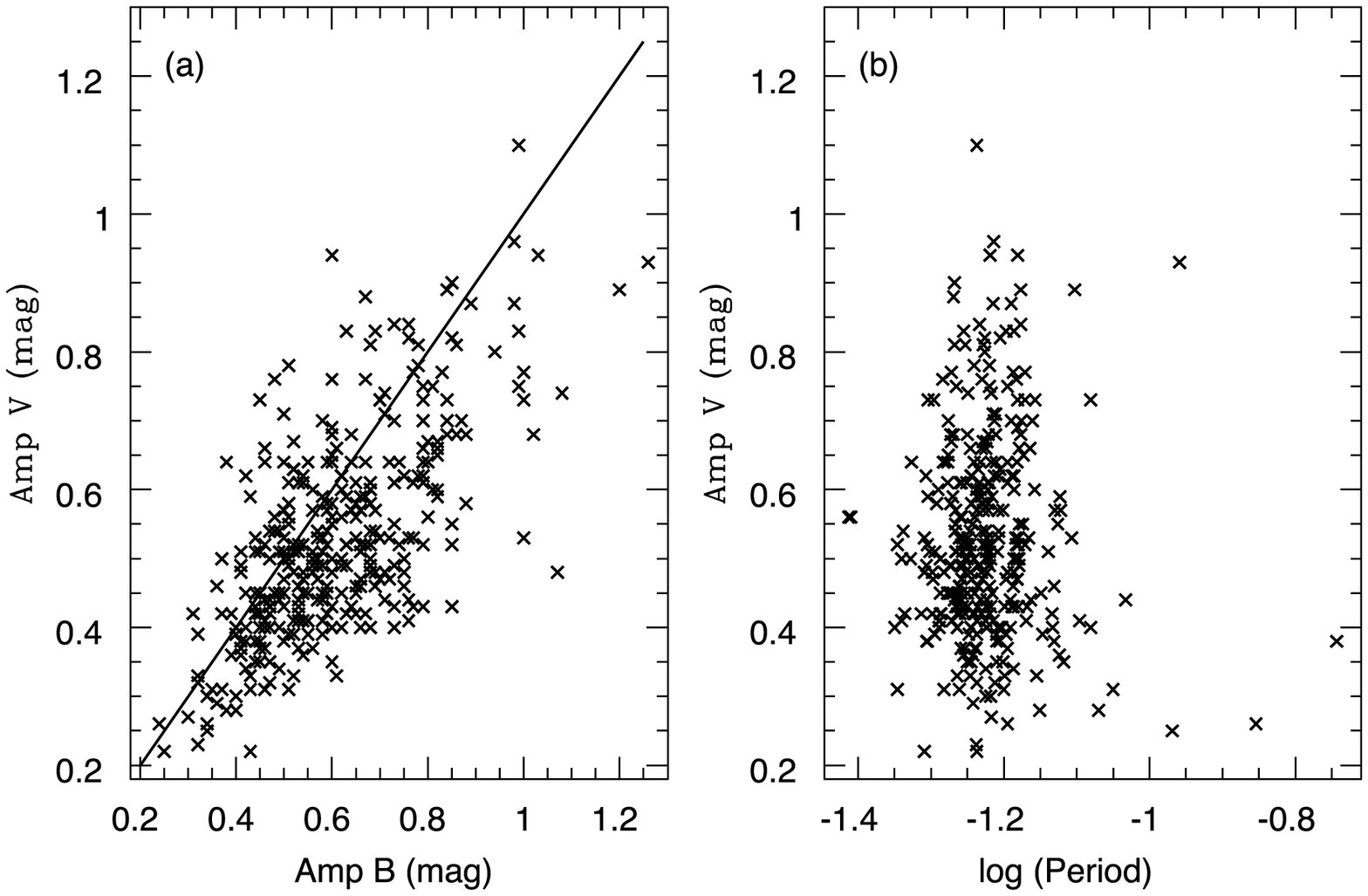}
 \caption{(a) Relationship between the amplitudes of DC stars in the B and V band. The solid
 line is a 1:1 relationship. 
 (b) Relationship between the amplitude in the V band and the logarithm of the
 period.}
 \label{fig-DCproperties}
\end{figure*}

All of the identified DC stars have relatively large amplitudes. 
The distribution of amplitudes in the B and V bands are shown 
in Figure~\ref{fig-DCamp}. In the B band, stars show variations between 0.2 and 1.6 magnitudes, 
with a mean value of 0.6 magnitudes. 
The V-band amplitudes range from 0.2 to 1.1 with a mean of 
0.5 magnitudes. However, these mean values may be biased toward larger mean amplitudes since it is
clear from Figure~\ref{fig-DCamp} (bottom) 
that we may be missing lower amplitude variables ($\lesssim 0.4$) 
at the faintest magnitudes. 
Figure~\ref{fig-DCproperties}a shows the relationships between the amplitudes in the B and V bands.
Most of the stars lie below the 1:1 line indicating that,
like other pulsating stars, amplitudes in the B band are 
larger than in V. The mean ratio between the amplitudes, Amp B/Amp V, is 1.2.

A period-amplitude diagram (Figure~\ref{fig-DCproperties}b) shows no correlation at all between 
these two properties of the light curve. It is well known that such a relationship exists for RR
Lyrae stars \citep[eg][]{smith95} with the lower amplitude variables having the longest periods.
However this seems not to be the case for DC variables.

Table~\ref{tab-DC} contains the properties of all the DC stars identified in this work.
The table includes the suspected alternative periods (aliases), and
the ID and periods determined by \citet{mateo98}. Stars within the locus of the fundamental
mode pulsators (see \S~\ref{sec-PL}) are identified with an "F" in column 13.

\subsection{An Eclipsing Binary and other periodic variables}

Out of the group of miscellaneous stars, 
Mis-5, shows a light curve typical of eclipsing systems (bottom left panel in Figure~\ref{fig-lc}). 
We measured
a period of 0.204 d for that star. The location of this star in the CMD is shown in Figure~\ref{fig-CMDf}.
It is separated from the main bulk of DC stars previously discussed.

The remaining 8 periodic variables in this group do not seem to fit in any of the groups
discussed above. They are marked with small open circles in Figure~\ref{fig-CMDf}. 
The periods of this
group of stars range 0.14 to 0.25 days, which are higher than what we found for the 
majority of DC stars (see Figure~\ref{fig-periods}). 
However, they do not occupy a clear region in the CMD.
They may be eclipsing systems with sinusoidal light curves.
Properties of these stars are listed in Table~\ref{tab-Mis}.

\section{Extinction}

\begin{figure*}[t]
\epsscale{0.9}
 \plotone{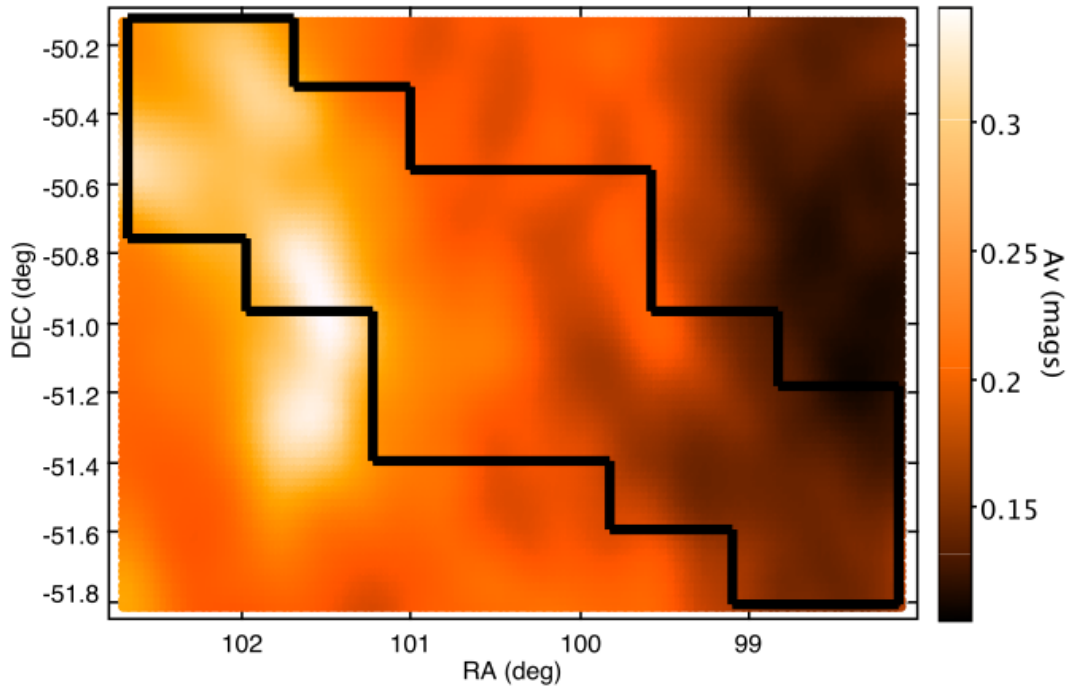}
 \caption{Extinction map from \citet{schlegel98} over the region of the survey. The 
 solid black lines enclose the observed region.}
 \label{fig-ext}
\end{figure*}

Since the sky coverage of this survey is large, we determined reddenings toward each of the
individual DC stars using the dust maps of 
\citet{schlegel98} instead of using a mean extinction 
for the whole galaxy. The mean galactic latitude of Carina
is $b\sim -22\fdg 2$, which is low enough to expect variations in extinction across the survey area.
Figure~\ref{fig-ext} shows a map of the $A_V$ extinction derived from 
the dust maps of \citet{schlegel98}\footnote{We used the tools available at the
NASA/ IPAC Infrared Science Archive \\
(\url{http://irsa.ipac.caltech.edu/applications/DUST/})}. It is clear from that 
map that there is a significant extinction gradient along the survey area, with the 
SW extreme of the survey having 
about twice the redening than the other extreme. In the central region of the galaxy the color excess, E(B-V), spreads over $\sim 0.01$ mags, which translates in 
differences in $A_V$ of the order of 0.03 mag. 
The mean value of the color excess of the DC stars 
is $\langle E(B-V) \rangle = 0.063$ mags with a standard deviation of 0.004 mags. Using
$A_V = 3.240 \times E(B-V)$ and $A_B = 4.325 \times E(B-V)$ \citep{schlegel98}, the mean 
extinctions are $\langle A_V\rangle = 0.20$ mags and  $\langle A_B\rangle = 0.27$ mags.

\section{Period-Luminosity Relationship \label{sec-PL}}

DC stars can be used as standard candles since they exhibit a period-luminosity (P-L) relationship.
A difficulty is that there are different relationships for stars pulsating in the 
fundamental (F) or first
overtone (FO) modes, with the latter being brighter for a given period. 
Thus, identification of the pulsation mode is  is crucial to assigning the proper P-L 
locus to a given star. 
Contrary to the RR Lyrae stars for which it is relatively easy 
to identify the pulsation mode based in the periods and shape of the lightcurves
(type ab and c), in DC stars the separation is more complex. Not only the periods
of both types of pulsators overlap but also, a "sawtooth" or sinusoidal light curves does not 
necessarily correspond with fundamental and first overtone pulsators 
\citep{mateo98,poretti08,cohen12}. Location in the CMD also does not help to separate stars
by mode \citep{cohen12}.

\begin{figure*}
\epsscale{0.9}
 \plotone{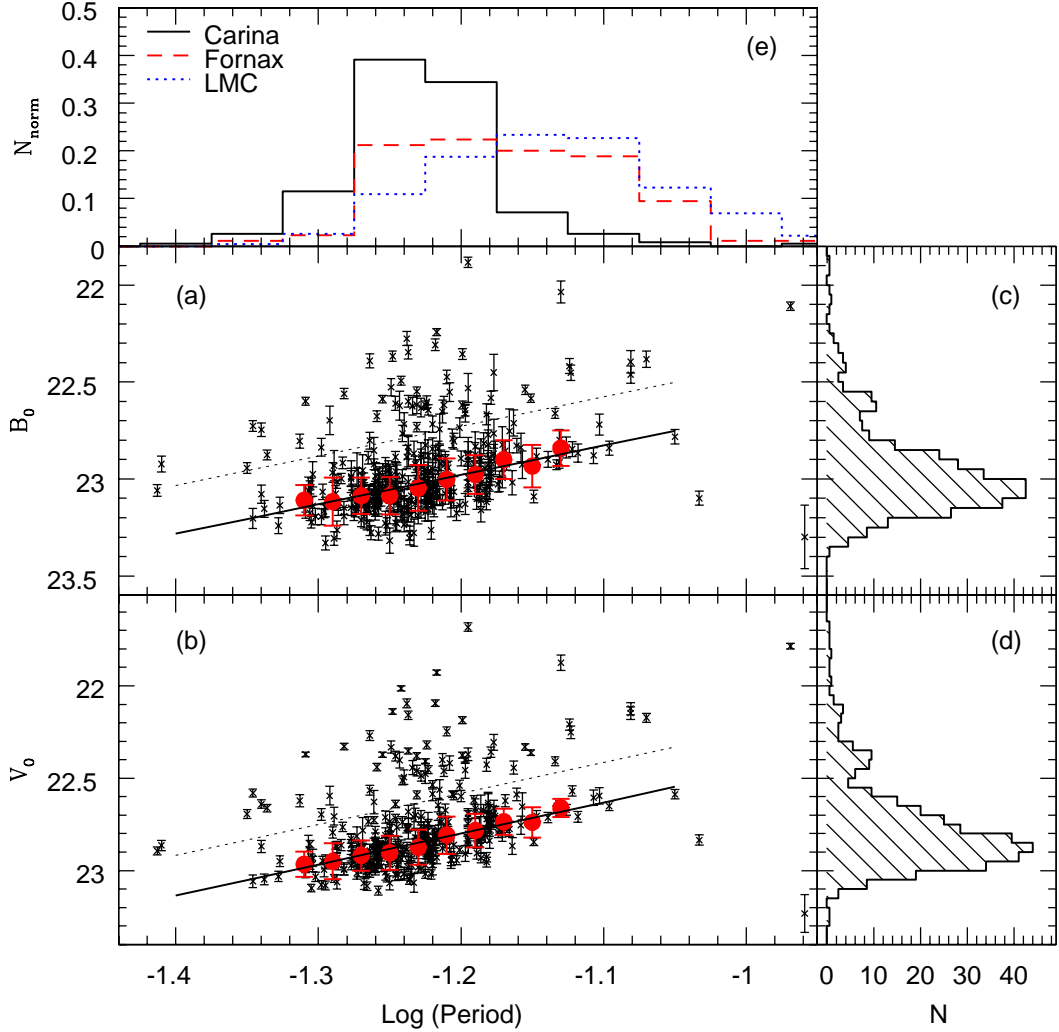}
 \caption{(a-b) Relationship between the extinction corrected magnitudes versus de logarithm of the 
 period of the 340 DCs found in this work ($\times$ symbols). 
 The stars pulsating in the fundamental mode are
 mostly located below the dotted line. 
 Large solid points are averages of the brightness of the fundamental mode pulsators 
 in bins of 0.02 in $log(P)$, and the error bars are the
 $1\sigma$ spread in each bin. Only bins with 5 or more stars were considered. The solid line is the 
 best fit through the red points. 
 (c-d) Distribution of the extinction corrected brightness in the B and V
  bands. 
  (e) Distribution of the periods of DCs in Carina, Fornax and LMC.}
 \label{fig-PL_histo}
\end{figure*}

Plots of the extinction corrected $B_0$ and $V_0$ magnitudes of the DC stars in Carina 
versus the logarithm of their periods (Figures~\ref{fig-PL_histo}(a) and \ref{fig-PL_histo}(b))
show several interesting things. First, it is clear that the DC stars in Carina show
a P-L relationship since fainter stars tend to have shorter
periods. Second, the distribution of periods of the DC stars in Carina
(panel e in Figure~\ref{fig-PL_histo}) has significant differences with the ones in 
the Fornax dwarf spheroidal galaxy and in the LMC; we defer a more detailed 
discussion about this issue for next section. 
Finally, the distribution of stars in this diagram is not uniform in magnitude.
Two groups or sequences of stars are separated by a gap containing very few stars. The fainter group is
the most numerous containing about 80\% of the whole sample. Then, a brighter group
seems to run parallel to the faint one but $\sim 0.35$ mags brighter. There is
a hint of a third sequence at even brighter magnitudes. This uneven distribution can 
also be seen in the histograms of magnitudes in the right panels
of Figure~\ref{fig-PL_histo} (panels c and d), in which two (maybe three) peaks are observed,
corresponding approximately with the sequences just described.
The behaviour is approximately the same in both the B and the V bands.

\begin{figure*}
\epsscale{0.9}
 \plotone{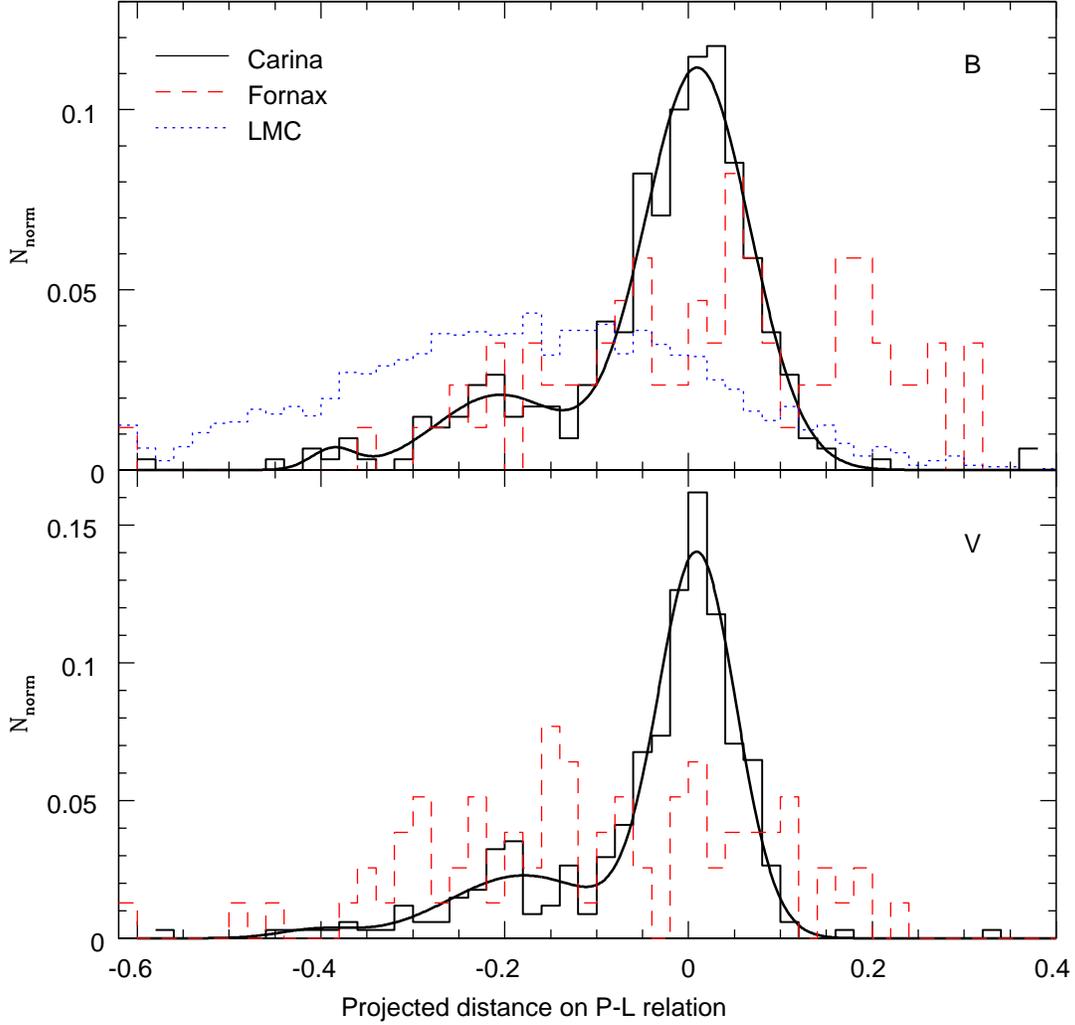}
 \caption{Histogram of the projected distance of all DC stars on the P-L relationship determined for Carina
 (equations~\ref{eq-PLB} and~\ref{eq-PLV}). The distribution of the projected distances
 in Carina is well modelled by three Gaussian curves (thick solid line), which we 
 interpreted as 
 groups of DC stars pulsating in different modes. For comparison, the
 distributions for Fornax and the LMC are also shown as dashed and dotted histograms.
 A distance modulus of 20.72 and 18.47 was assumed for Fornax and the LMC respectively 
 in order to calculate the relative distance on the P-L relationship of Carina. 
 All histogram have been normalized by the total number of DCs in each galaxy.}
 \label{fig-dif}
\end{figure*}

To make a more quantitative assessment of the distribution of the stars in this diagram, 
we drew an arbitrary line through
the gap and fitted a straight line to the most numerous group, that is, all the stars below 
the line. Then, we calculated the projected distance of each star to the best fitted line
(Figure~\ref{fig-dif}). A histogram of the distribution of the projected distances in the B and 
V band clearly shows that the uneven distribution is well modeled by three Gaussian 
curves corresponding to the sequences described above. 
The smallest Gaussian corresponding to the brightest sequence is better 
seen in the B data, and just barely in V.
The minimum point between the two main Gaussian curves in Figure~\ref{fig-dif} 
allowed us to better define the
separating line between the two strongest sequences of stars in the diagram. 
After a few iterations in this process we found the best separation of the sequences,
which is shown as a dotted line in Figure~\ref{fig-PL_histo}. 

With only 20 DC stars in Carina available at that time, \citet{mateo98} was already able to distinguish the
two strongest sequences observed in Figure~\ref{fig-PL_histo}, and recognized the faintest one
as stars pulsating in the F mode, while the brightest sequence
corresponded to FO pulsators. This was made by realizing that the PL relationships
for these two types of pulsators, as given by \citet{nemec94}, fitted very well their data and 
implied a distance modulus for Carina of $\mu_o=20.06$, in good agreement with
other measurements in the literature. 
Assuming this distance, the \citet{nemec94}'s PL relationships (which include a metallicity 
term for which we assumed, as \citeauthor{mateo98} did, [Fe/H]$=-2.0$) 
pass over the two sequences observed in our data as well, confirming that the bulk
of the DC stars in Carina should be stars that are pulsating in the F mode, 
and the parallel brighter
sequence corresponds to stars pulsating in the FO. The brightest stars in the sample
may be pulsating in even higher overtones. 

We thus selected F pulsators as the stars below the arbitrary line shown in Figure~\ref{fig-PL_histo}.
Because we see such a clear distinction among the DCs in Carina, we can readily investigate
 the pulsational properties of the DCs as a function of mode.  
 For example, upon examination
of some of the light curves of stars pulsating in different modes, we arrived at the
same conclusion as previous authors \citep{mateo98,poretti08} that the shape of the light 
curve is not a indication of the pulsation mode. The third row of lightcurves in 
Figure~\ref{fig-lc} shows three examples of DC stars in our sample. The middle and right 
column (DC-262 and DC-197)
show two stars in the F pulsation locus. One of them has a clear sawtooth-type lightcurve
while the other is almost sinusoidal. The star in the left column 
(DC-87) is a FO pulsator according to its location
in the PL diagram. It has a sawtooth lightcurve. There are more examples of similar cases 
in our Carina dataset.

To determine the P-L relationship for the Carina DCs, we calculated the average brightness of 
F pulsators in 0.02 dex bins in $\log(P)$ that contained five or more stars; these binned 
averages are shown as large dots in Figure~\ref{fig-PL_histo}.   
This method gives equal weight along the full period range while minimizing the effects of 
outliers.  We then use a least squares method to fit a straight line through those points (solid 
lines). We found the following relationships:

\begin{equation}
B_0 = 21.16 - 1.520 \log(P)
\label{eq-PLB}
\end{equation}

\begin{equation}
V_0 = 20.78 - 1.682 \log(P)
\label{eq-PLV}
\end{equation}

As mentioned before, the P-L relationships given by \citet{nemec94} fitted well the earlier 
data of DC stars in Carina. However, the best fit to these new Carina data 
has a shallower slope than the P-L
relationship in \citet{nemec94}, which for the F mode pulsators in the V band is:

\begin{equation}
	\begin{array}{ll}
		M_V= -2.56 (\pm 0.54) \log(P) + 0.32 {\rm [Fe/H]} + 0.36; \ \;	& \mbox{(Nemec et al 1994)} \\
	\end{array} 
\label{eq-N94}
\end{equation}
 
In recent years there have been several efforts in establishing precise P-L relationships for
DC stars which have even steeper slopes than \citet{nemec94}. 
If the metallicity of the population is known, \citet{mcnamara11} has proposed a
metallicity-dependent P-L relationship for DC stars:

\begin{equation}
	\begin{array}{ll}
		M_V= -2.90 (\pm 0.05) \log(P) -0.19 {\rm [Fe/H]} -1.27; \ \;	& \mbox{if {\rm [Fe/H]}$>-2.0$}  \\
		M_V= -2.90 (\pm 0.05) \log(P)-0.89; &  \mbox{if {\rm [Fe/H]} $\leqslant -2.0$} 
	\end{array} 
	\label{eq-Mc11}
\end{equation}

In practice, for extragalactic systems it is typically not straightforward to define a single 
metallicity for purposes of defining the P-L relation for a DC population (and especially since 
both SX Phe and $\delta$ Scuti
stars may coexist in the same place).  
\citet{poretti08} and \citet{cohen12} have constructed P-L relationships
by joining datasets from galactic globular cluster and some extragalactic systems (namely, 
Fornax, Carina
and the LMC; the latter included only in \citet{cohen12}'s relationship). The resulting P-L 
equations are:

\begin{equation}
	\begin{array}{ll}
		M_V= -3.65 (\pm 0.07) \log(P) -1.83; \ \;	& \mbox{(Poretti et al 2008)} 
	\end{array} 
	\label{eq-P08}
\end{equation}

\begin{equation}
	\begin{array}{ll}
		M_V= -3.389 (\pm 0.09) \log(P) -1.640; \ \;	& \mbox{(Cohen \& Sarajedini 2012)} 
	\end{array} 
	\label{eq-C12}
\end{equation}

We applied all those P-L relationships to our data to calculate the distance 
modulus to Carina using its 268 F-mode DC pulsators by
determining the best-fit for the fixed
slopes of relations~\ref{eq-N94}  to~\ref{eq-C12}.
From the intercept of the best-fit lines in combination
with the equations above we derived a de-reddened distance modulus for Carina.
Table~\ref{tab-distance} summarizes the results. For the
relationships that have a dependency on metallicity, we used both [Fe/H]$=-2.0$ and $-1.7$,
although the latter may be more representative of the intermediate/old population of Carina 
\citep{monelli03}. For [Fe/H]=$-1.7$, the average of the distance modulus given by 
equations~\ref{eq-N94}  to~\ref{eq-C12} is $\mu_0=20.17 \pm 0.10$ mags, which is in very 
good agreement with the estimate from RR Lyrae stars by D03 ($\mu^{RR}_0=20.12 \pm 0.12$).

\section{Comparison with the Fornax Dwarf Spheroidal Galaxy and the LMC}

Numerous DC stars have also been found in the Fornax dwarf spheroidal galaxy by \citet{poretti08}.
These stars in Fornax are referred as SX Phe by \citet{poretti08} since they expect most of the
stars in this galaxy to be metal-poor.
Like Carina, Fornax contains a complex stellar population with a prominent intermediate age population 
\citep{coleman08}
and a well populated main sequence at the intersection with the instability strip.
\citet{garg10} identified 2,323 DC stars (referred in this case as HADS) 
in the LMC using 
the SuperMACHO dataset. 
The number of DC stars in Carina, Fornax and the LMC are large enough to allow detailed 
comparisons on their properties.
Three SX Phe have been reported in the globular cluster M54 located at the center of the
Sagittarius dSph galaxy \citep{sollima10} 
but they are too few to make any firm comparison of their properties.
To our knowledge, these are the only extragalactic DC systems known.

\begin{figure*}[t]
\epsscale{0.7}
 \plotone{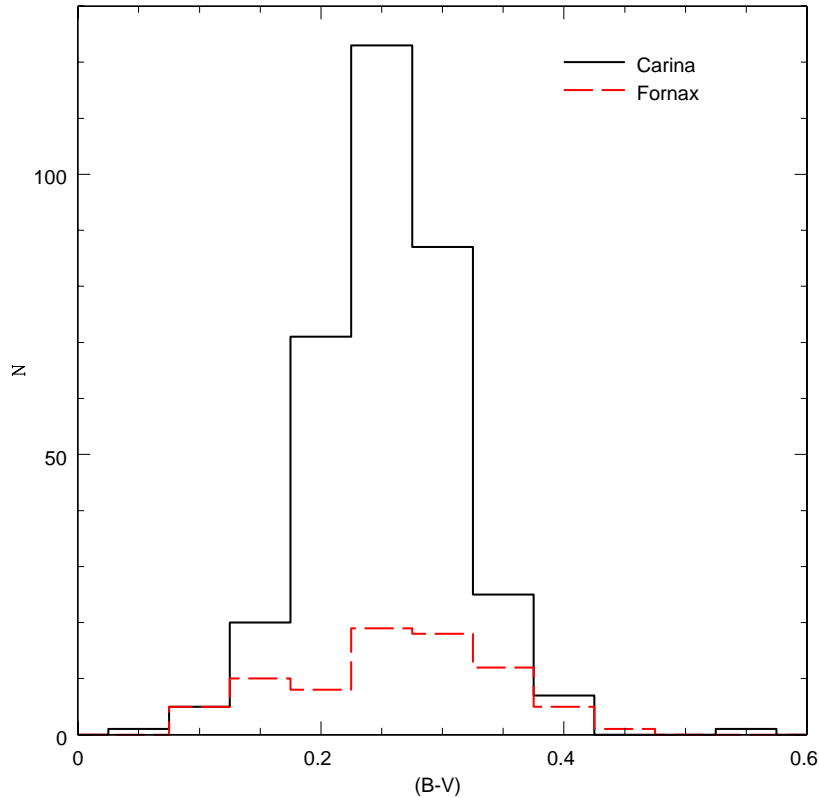}
 \caption{Color distributions of DC stars in the Carina (solid histogram) and 
 Fornax (dashed histogram) dwarf spheroidal galaxies. }
 \label{fig-color}
\end{figure*}

Based on the color distribution of the DC stars (Figure~\ref{fig-color}) we find that the width of the
instability strip is remarkably similar in both Carina and Fornax. 
In both cases, most of the DC stars are confined
to the color interval $0.10 < B-V < 0.40$. \citet{poretti08} make the case, however, that
this width may be larger than what is observed in the Milky Way. We could not compare with the color
distribution in the LMC since the only color provided by \citet{garg10} is $B-I$.

The DC stars in the three galaxies have very short periods ($<0.1$ d).  
However, Figure~\ref{fig-PL_histo} (panel e) shows that the period
distribution of the Carina DCs is the shortest of the three. There seems to be a sequence among the three galaxies with the
distribution of periods shifted toward longer periods when going from Carina, to Fornax
and to the LMC.
The median value of
the periods in Carina is 0.059 days, while in Fornax is 0.067 days and in the LMC is 0.073 days. We notice that the mean metallicity in these galaxies has the same sequence with Carina being the most metal-poor of the three. 

Aliasing should not be responsible for the unequal period distributions since
the differences between the true periods and their most common aliases are very small,
much smaller than the bin size of the histograms in Figure~\ref{fig-PL_histo} for stars with
periods $<0.1$ d. On the other hand, it is possible that incompleteness may play a role, at least 
in part, for the differences observed in the distributions.
Stars with the shortest periods are also the faintest ones (they follow a P-L relationship). 
In the case of Fornax (as in Carina) the limits
on the minimum amplitude for the detection of variables increases with magnitude (see our
Figure~\ref{fig-DCamp} and Figure 2 in \citet{poretti08}). Hence, stars with small amplitudes
are likely missing at the faintest magnitudes. The minimum amplitude observed in 
DC variables in these two galaxies varies as a function of magnitude approximately in 
the same way within the range of magnitudes of the DC stars. Thus, incompleteness should 
affect both galaxies similarly, while for the LMC incompleteness may not be significant since 
the minimum amplitude detected remains constant as a function of magnitude, at least 
in the range of interest of these stars. 

Both Fornax and the LMC have a larger relative number of stars with periods $>0.075$ d,
which are rare in Carina. 
Similarly, but not as striking, the LMC has a relatively large number of stars
with periods $>0.095$ d, which are rare in Fornax.
Kolmogorov-Smirnov tests reject the hypothesis that the samples are drawn from the 
same parent population at the 99.9\% confidence level. 
The longer periods among the DC stars in the LMC can be 
understood if this galaxy contains mostly $\delta$ Scuti stars (specifically, HADS), which is indeed the 
identification given by \citet{garg10} to those stars. HADS stars tend to have longer periods than SX Phe,
although significant overlap in the period distributions of both types exists \citep{mcnamara11}.

\begin{figure}
\epsscale{0.9}
 \plotone{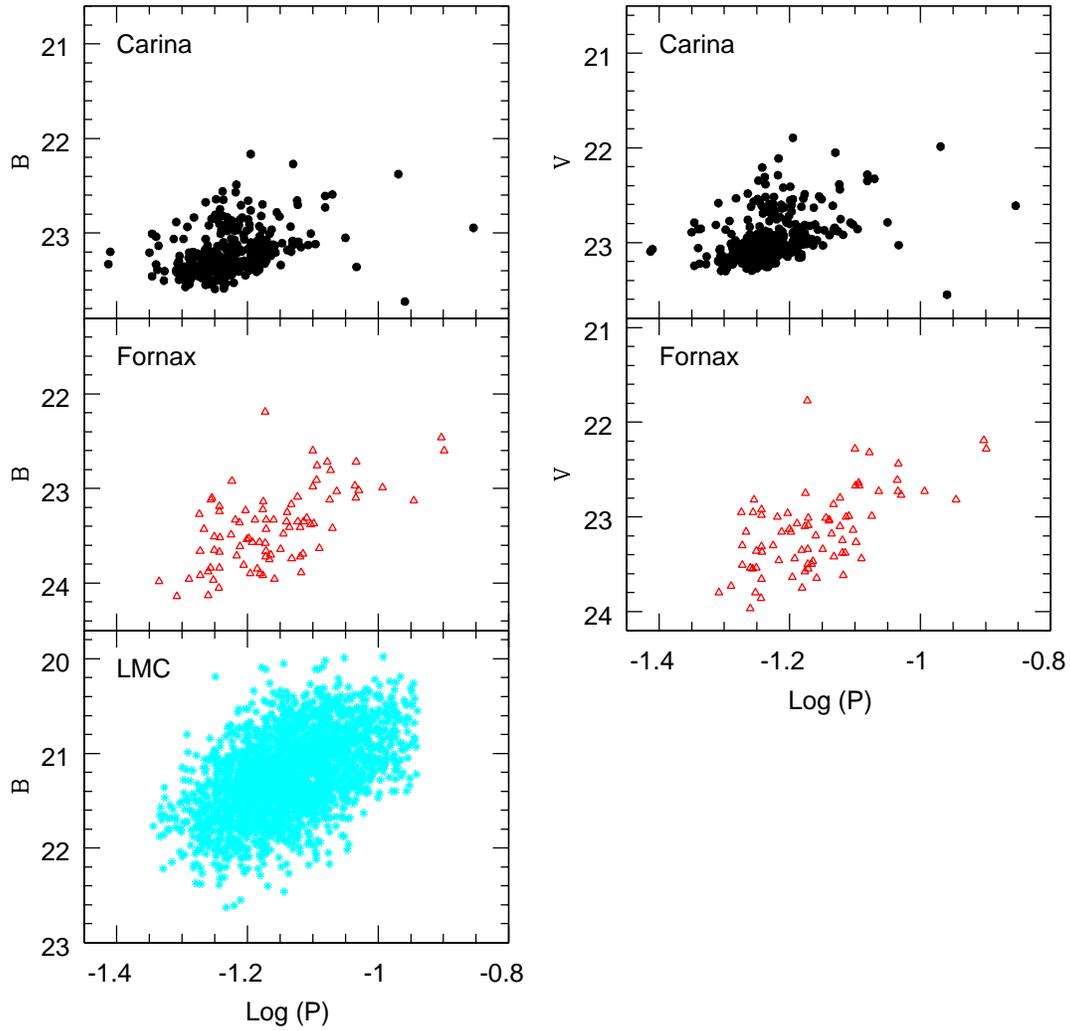}
 \caption{Observed period-luminosity diagrams in the B and V bands for Carina (top panels), 
 Fornax (middle panels) and the LMC (lower panel). Data for Fornax and the LMC are from
 \citet{poretti08} and \citet{garg10}, respectively. For a better comparison,
 the vertical axis scale in all panels is 
 the same.}
 \label{fig-PL_CFL}
\end{figure}

Another remarkable difference in the population of DC stars among these three galaxies has to 
do with the distribution of stars in the P-L diagram. 
As mentioned in section~\ref{sec-PL}, the 
stars pulsating in the F mode in Carina are clearly separated in the diagram by a gap
containing almost no stars. However, that is not the case for Fornax and the LMC. 
In Figure~\ref{fig-PL_CFL} we show the P-L diagrams for the three galaxies. Fornax and the LMC show a wider distribution (in both periods and brightness) with no obvious gaps or 
distinct sequences apparent. This is more clearly seen in Figure~\ref{fig-dif} in which we 
plotted the projected distance of the stars in the three systems on the P-L relationship
determined for Carina (equations~\ref{eq-PLB} and~\ref{eq-PLV}), after taking into account 
the difference in distance modulus of those galaxies with respect to the value of 20.17 
we found in this work for Carina (we assumed a distance modulus of 20.72 for Fornax 
\citep{rizzi07} and 18.47 for the LMC \citep{cohen12}). As we explained in the previous 
section, in Carina this distribution is well modeled with three Gaussian curves which we
interpreted as sequences of stars pulsating in different modes. This is not observed 
in Fornax or in Carina since both galaxies present a more flattened distribution of 
projected distances on the P-L relationship.  No clear peaks are present in the histograms
for these galaxies.

A spread in metallicity and/or a significant depth along the line of sight may be possible
explanations for these differences. 
The contribution of depth along the line of sight, however, should have
only a small contribution in the observed spread in brightness, at least in the case of Fornax. 
A spread of 5 kpc along the line of sight in Fornax
(which is about its angular size) should produce a spread in magnitude of only 0.02 mags 
at the 
distance of this galaxy (120 kpc). This is too small to be responsible of filling up the gap 
between the two sequences of pulsators. 
A spread in metallicity could have a larger effect. From 
equations~\ref{eq-N94} and~\ref{eq-Mc11}, it is inferred that a spread of $1.0$ dex in [Fe/H] would
produce variations in absolute magnitude of the order of $0.2-0.3$ mags, which are large enough
to account for the spread observed in Fornax and the LMC. Spectroscopy of Fornax shows
that the galaxy has a significant metallicity gradient \citep{pont04},
which is probably not the case for Carina. \citet{bono10}
argued that the intermediate age population of Carina has only a modest metallicity spread of  
$\sim 0.5$ dex at most.

A third difference among the DC population in these three galaxies is the slope of the PL relationship.
The vertical axis in Figure~\ref{fig-PL_CFL} has the same scale for the three galaxies. It is clear that 
the PL relationship in both Fornax and the LMC have a steeper slope than Carina. Indeed, \citet{poretti08}
fits the F pulsators in Fornax with a slope of $-3.3$ (in the V band), while \citet{garg10} found a slope of
$-3.43$ in the LMC (in the broadband VR filter). The slope we find in Carina ($-1.68$ in V) is
significantly shallower. Part of the difference may be due to the fact that
the narrow range of periods found among the Carina DC stars may not 
provide a strong constraint on the slope of the PL relationship. But the effect of age and 
metallicity spreads within each galaxy on the P-L relationship slope remains to be studied.

From these comparisons it is clear that the DC population in these three systems show differences that
may be important to understand the evolution and star formation in these galaxies. All three
are known to have a complex mix of stellar populations, but their star formation histories are likely quite distinct.
Carina has produced a large number of DC stars, but they have narrower properties that may 
indicate that they have a common origin. Presumably, most of them are main sequence stars from the 
populous intermediate-age population known to exist in this galaxy \citep{monelli03}. On the other hand,
Fornax, which also has a prominent intermediate-age population, has DC stars with broader properties
(both in period and luminosity). A significant spread of metallicities in the intermediate age population
of Fornax may contribute to their different mean properties. 
Another possibility, which was suggested by \citet{poretti08}
is that the DC sample in Fornax contains a mixture of stars arising from different evolutionary paths. 
Blue stragglers from the old population in that galaxy may 
also populate the instability strip in this galaxy.
However, Carina also has an old population (it has numerous RR Lyrae stars, as shown above), 
and thus it may also contain, in principle, a combination of main
sequence and blue stragglers among its DC stars. If so, the properties of both type of stars have 
very
similar properties in Carina, which is not the case for Fornax. Alternatively, the larger spread 
observed among the Fornax and LMC DC stars may be a reflection of the contribution of old stars to the 
total population in these galaxies. CMD studies show that in Carina the old population may be as 
little as $10\%$ \citep{hurley98}, while it is significantly higher in Fornax and the LMC 
\citep{mateo98rev,grebel99,coleman08}.
Detailed comparisons with stellar 
evolutionary models may help to understand the origin and properties of DC stars under different
conditions.

\section{The Extended Spatial Distribution of the Variable Stars in Carina}

Figure~\ref{fig-sky} shows that there are RR Lyrae stars,
anomalous Cepheids and DCs well outside the \citet{king62} tidal radius of Carina
\citep{irwin95}.
The existence of these stars is additional proof of the tidal disruption of Carina. 
We detected DC stars up to a distance of $55\arcmin$ of the center of Carina. Similarly, RR 
Lyrae
stars and anomalous Cepheids  were detected up to $65\arcmin$ and $75\arcmin$ 
respectively.
The presence of these stars at such large distances is consistent with the results described
in \citet{battaglia12} in which both the old and intermediate age population of 
Carina extends to distances of $\sim 60\arcmin$. It would be very 
unlikely that these stars were actually field stars in the Milky Way since very few halo stars 
are expected at such large distances from the galactic center. 

\begin{figure}
 \epsscale{0.9}
 \plotone{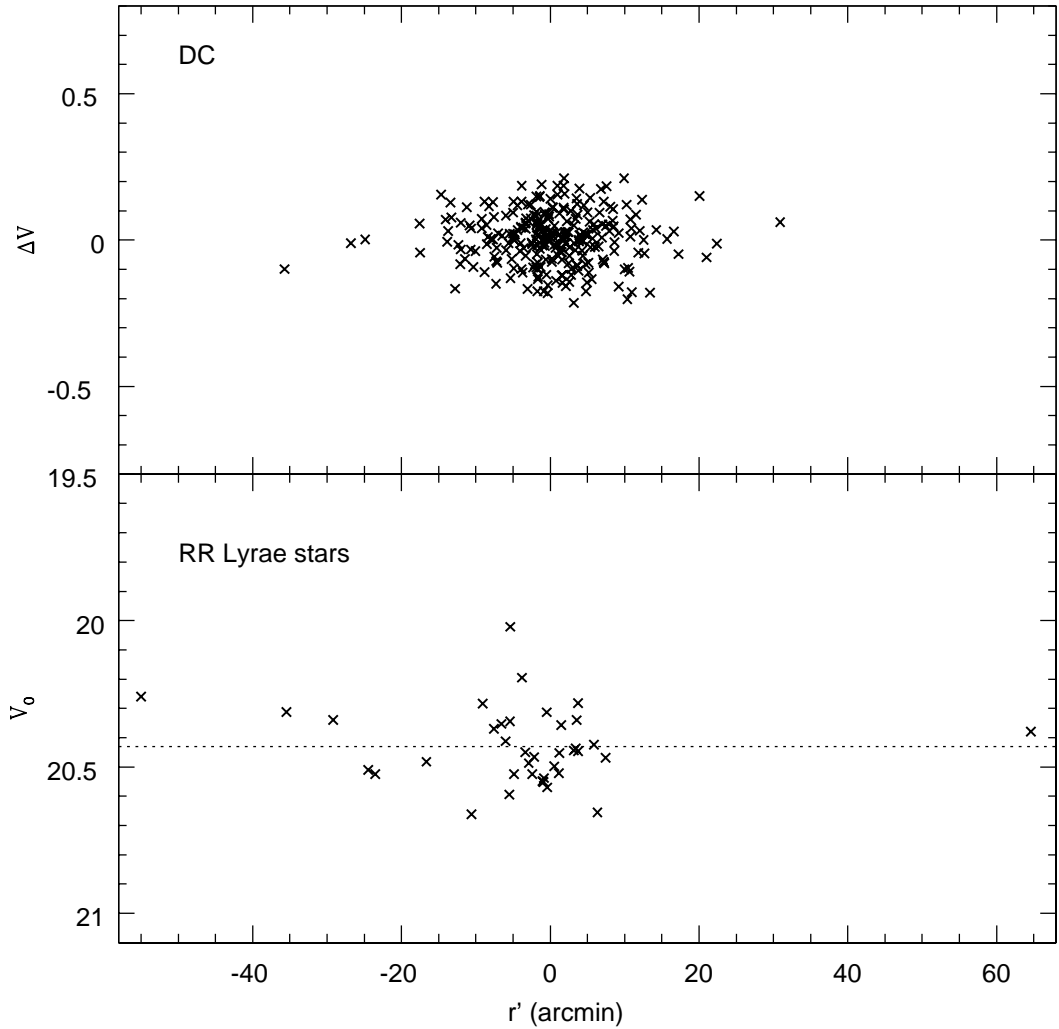}
 \caption{{\sl (Top):} Variation of $\Delta V = V_0-V_F$ along $r'$, the projected distance on the 
 semi-major axis, for all DC stars identified as F pulsators. {\sl (Bottom):} Variation of $V_0$ along
 $r'$ for all RR Lyrae stars. The dotted line indicates the value of the mean reddening-corrected $V$ magnitude of all the RR Lyrae.}
 \label{fig-deltaV}
\end{figure}

Since both DC and RR Lyrae stars are standard candles and they both extend to large 
distances from
the center of Carina, it is natural to ask if there is a gradient in distance from one side of the
galaxy to the other. Indeed, the best P-L relationship fitted to our data (Fig~\ref{fig-PL_histo}) shows
some dispersion which could be due, at least in part, to differences in depth along the line of sight
 of the DC stars. In order to investigate this issue, we calculated $\Delta V = V_0-V_F$ for all 
 the DC F-pulsators, where $V_F$ is the magnitude predicted by the P-L relationship fitted to the data
 (eq.~\ref{eq-PLV}). In the top panel of figure~\ref{fig-deltaV}, we show $\Delta V$ as a 
 function of $r'$, the
 projected distance of each star on the semi-major axis of the galaxy 
 \citep[assuming a position angle of $65\degr$,][]{irwin95}. In this plot, negative values of $r'$
 refer to the N-W side of Carina. It is clear that there is no trend of distance with position in the 
 galaxy. A similar conclusion is reached by looking at the variation of the magnitude of the
 RR Lyrae stars as a function of $r'$ (bottom panel in Figure~\ref{fig-deltaV}). No gradient is evident 
 in the extinction corrected V magnitude of the RR Lyraes stars along their projected distance
 on the semi-major axis of the galaxy.
 
 The dispersion of the P-L relationship of the DC in the V band (eq~\ref{eq-PLV}) is 0.09 mags. 
 This implies a 
 fractional error in distance ($\sigma_d/d$) of only 4\%. However, at the distance of Carina, a 
 fractional error of 4\% implies $\sigma_d\sim 4$ kpc. The projected size of Carina is only 1.3 kpc and thus it 
 would have been impossible to measure any internal sub-structure in the galaxy using these 
 stars unless Carina were very much more extended along the light of sight than in the transverse 
 direction. 
 The dispersion observed in the P-L relationship of the DC puts an upper limit in the size 
 of a gradient between the more distant DC stars detected on both sides of the galaxy, 
 $\Delta d/\Delta r'$. 
 If such gradient exists it should not be larger than 4 kpc$/67\arcmin = 0.06$ kpc/$\arcmin$.
A clear limitation of this conclusion is of course the small number of stars at large $|r'|$.

\section{The frequency of the Dwarf Cepheids}

In order to estimate the frequency of DC stars in Carina, we selected the central region of the galaxy
in which we have the overlap region of Fields 3 and 4 ($100.25 < \alpha (\degr) < 100.53$,
$-51.17 < \delta (\degr) < -50.77$). This region has also a large number of epochs
(see Figure~\ref{fig-Nobs}) and hence our completeness is likely to be very high in this region.
We selected stars in the upper part of the main sequence, namely stars within the 
following boundaries: $22.5 < V < 23.3$ and $0.15 < (B-V) < 0.35$. There are 2,531 stars
in this part of the CMD and 197, or $8\%$ of them, are DC stars.
This figure is higher than the one provided by \citet{mateo98}, 2\%.
However, as in \citet{mateo98}, our figure of $8\%$ should also be considered a lower 
limit in the number of DC stars produced in Carina 
since most likely we are missing small amplitude variables.
With a much lower detection threshold for the amplitude of the variables 
\citet{balona11} argue that no more than 50\% of the main sequence stars in the
Kepler field are pulsating.
In contrast, the frequency of pulsation at the 
RR Lyrae luminosities is nearly 100\%, and appears to 
be similarly high for anomalous Cepheids and classical Cepheids as well. 
The reason why not all the stars within the instability strip are DC stars (or at least DC stars with
amplitudes greater than 0.2-0.4 mags) remains to be explained.
Something about main sequence stars is able to suppress pulsation (surface convection, 
magnetic fields, rotation, higher surface gravity?) which is less effective or absent in more 
luminous pulsators. 

Results from the Kepler mission suggest that DC of high amplitude are extremely rare among
field stars \citep{balona12}. Indeed, only one out of more than 1600 DC stars identified by
Kepler varies with an amplitude $>0.3$ mags. This finding suggests two possible scenarios: either 
the high amplitude DC found in the extragalactic systems are just the tip of the iceberg of
a much larger population which is below detection with current observations, or the properties
of DCs in the galactic field are substantially different from the ones in extragalactic systems.
The number of high amplitude DC stars in Carina suggests that the later is a more feasible option.
If the frequency observed in the field were to hold in Carina, we would expect that for the 197 high amplitude
DC in the central part of Carina, there should be $\sim 300,000$ low amplitude pulsators, which is $\sim 10^2$ 
times the total number of stars detected in that part of the CMD. Thus, the frequency of high
amplitude DC stars in Carina is at least $100\times$ higher than in the field.

\subsection{Specific frequency of DC in Local group galaxies and Galactic globular clusters}

We calculated the specific frequency of DC stars, $S_{\rm DC}$, by
adapting the definition used by \citet{mateo95} for the specific frequency of anomalous 
Cepheids, that is, $S_{\rm DC} = N_{\rm DC} / L_V$, with the luminosity in units of $10^5 
L_\odot$. In the case of Carina we corrected the number of DC stars to include only stars inside 
the tidal radius \citep[as given by][]{walcher03}. In the case of Fornax, we
corrected the number of DC for the fact that the field observed by \citet{poretti08} did not 
cover the whole galaxy. We calculated the observed fraction of the luminosity of Fornax by
integrating the \citet{king62} profile \citep[see][]{mateo95,vivas06}
with the parameters $r_c=13\farcm 7$ and $r_t=71\farcm 1$
\citep{irwin95}. The ellipticity and position angle of the galaxy were taken into account to
determined the observed fraction of the galaxy, which turned out to be 0.88. Thus, the
estimated number of DC in Fornax is $85/0.88=97$. We did not attempt to correct the
number of DC in the LMC given the complexity of the superMACHO dataset \citep{rest05}.
However, that survey covers the entire central part of the LMC (hence, most of the light), and
the specific frequency we calculated is likely close to the real value. Still, it should be 
considered just as an approximate value.  

DC stars have been detected as well in several Galactic globular clusters. We selected the clusters
with $\geq 5$ DC stars from the Catalog of Variable Stars in Globular Clusters 
by \citet{clement01}. For the globular clusters M55 and NGC5466, we took the 
updated (June 2013) value of $N_{\rm DC}$ in Clement et al's 
catalog\footnote{\url{http://www.astro.utoronto.ca/{\tiny$\sim$}cclement/read.html}},
which includes more recent observations of those clusters by \citet{pych01} and \citet{jeon04}.

The specific frequencies of DC for all these stellar systems
are reported in Table~\ref{tab-frequency}.
We did not make any consideration for missing variables in any of these stellar systems. If 
incompleteness were an important issue, the specific frequencies would be higher than 
the ones we calculated here.
Carina is the stellar system with the highest specific frequency of DC stars known to date.
It has a significant higher specific frequency than the Fornax dSph galaxy and the LMC. 
It also have more DC stars (per luminosity) than all the globular clusters than have been searched for this type of variables.

\section{CONCLUSIONS}

We have searched the Carina dSph galaxy for DC stars, extending significantly the area 
first explored by \citet{mateo98}. With 340 DC stars, 
we find that Carina is very rich in this type of
stars which is likely linked to the fact that this galaxy has a significant intermediate age 
population
and hence, a well populated main sequence in the region of overlap with the instability strip.
It has the highest specific frequency of DC among the Local Group galaxies and galactic globular 
clusters that have been searched for this type of stars. 

The DC stars in Carina spread from 1.2 to almost 3 magnitudes below the horizontal branch.
Most of the stars (80\%), however, are faint and clump around $V\sim 23.1$ (or $\sim 2.5$ mags 
below the horizontal branch).  These stars are identified with ones pulsating in the fundamental
mode, which follow a tight P-L relationship. Using several calibrations of the P-L relationship for
DC stars, we derived a distance modulus of $\mu_0=20.17 \pm 0.10$ which agrees very well
with the distance derived by D03 using RR Lyrae stars.

The survey covers the whole galaxy and extends an angular distance of $\sim 100\arcmin$ 
from the center of Carina along the semi-major axis, in both directions.
Although our completeness for DC stars
is low and not uniform in the most external parts of the survey, we discovered several DC
stars located up to $\sim 1\degr$ from the center which is
equivalent to $\sim 2-3$ times the tidal radius of Carina (depending on the assumed tidal radius
which vary from $22\arcmin 5$ \citep{walcher03} to $28\arcmin$ \citep{irwin95}).  
The existence of these extra-tidal stars in the Carina field is reinforced as well by the discovery of
RR Lyrae stars and anomalous Cepheids at large distances from the center, and
confirms previous findings that this galaxy is under tidal disruption. In particular, the fact that
all three types of variables are found far away from the center confirms that both the old
and intermediate age population in Carina extends beyond its tidal radius 
\citep[see][]{battaglia12}.

The properties of DC stars in dSph galaxies reflect the different star formation histories of these
galaxies and possibly their mean metallicities. 
Carina's DC stars differ significantly in several aspects with respect to the ones in Fornax and 
the LMC. The period distribution in Carina is narrower than
in those other two galaxies. Also, the mean period in Carina is smaller. Only in Carina the 
sequence of fundamental pulsators is well separated from the first overtone one. 
We speculate that most of the
DC stars in Carina come from a stellar population which has only a small spread in metallicity.
In Fornax a large range of metallicity for the intermediate age population may be the 
reason for the larger spread observed in the diagram of brightness versus period. 
This possibility has grounds in the study of the populations of those galaxies with
other methods.  A detailed analysis of the CMD of Carina \citep{bono10} and
spectroscopy in Fornax \citep{pont04} indicate that the spread in metallicity in Carina is probably 
very small, while that is not the case for Fornax.
In the LMC, depth
along the line of sight may also play a role in the observed properties of the DC stars. 
As in Carina,
Fornax is also too far away for depth effects to likely have a measurable effect in the PL diagram.

Our findings in Carina suggest that DC stars are a good tool to find tidal tails of disrupting 
systems. They should be useful as well to find streams in the Milky Way's halo if they
have an intermediate-age population associated with them.
In Carina, DCs are $\sim 3-4$ times more numerous than RR Lyrae stars, although that ratio may
be different in other systems depending on their particular combination of stellar populations.

Further observations of DC stars in other extragalactic systems and comparison with existing and 
updated theoretical models of stellar evolution and stellar pulsation will help
to shed more light on their origin and their
production under different conditions of age and chemical composition.

\acknowledgments
Based on observations obtained at the Cerro Tololo Inter-American Observatory, National
Optical Astronomical Observatories, operated by the Association of Universities for Research
in Astronomy (AURA) under cooperative agreement with the National Science Foundation.
AKV thanks the hospitality of the Department of Astronomy at University of Michigan 
during her sabbatical leave in which most of this work was made.

{\it Facility:} \facility{Blanco}

\appendix 

\section{Lightcurves \label{ap-lc}}

The full set of light curves are presented as online-only material in Figures 20.1-20.4
(RR Lyrae stars), 21.1-21.29 (DC stars), 22.1 (anomalous Cepheids), and 23.1 (Miscellaneous
periodic variable stars).
 
\figsetstart
\figsetnum{20}
\figsettitle{RR Lyrae Stars}

\figsetgrpstart
\figsetgrpnum{20.1}
\figsetgrptitle{Lightcurves of RR-1 to RR-12}
\figsetplot{f20_1.eps}
\figsetgrpnote{Lightcurves of RR Lyrae stars}
\figsetgrpend

\figsetgrpstart
\figsetgrpnum{20.2}
\figsetgrptitle{Lightcurves of RR-13 to RR-24}
\figsetplot{f20_2.eps}
\figsetgrpnote{Lightcurves of RR Lyrae stars}
\figsetgrpend

\figsetgrpstart
\figsetgrpnum{20.3}
\figsetgrptitle{Lightcurves of RR-25 to RR-36}
\figsetplot{f20_3.eps}
\figsetgrpnote{Lightcurves of RR Lyrae stars}
\figsetgrpend

\figsetgrpstart
\figsetgrpnum{20.4}
\figsetgrptitle{Lightcurves of RR-37 to RR-38}
\figsetplot{f20_4.eps}
\figsetgrpnote{Lightcurves of RR Lyrae stars}
\figsetgrpend

\figsetend

\figsetstart
\figsetnum{21}
\figsettitle{Dwarf Cepheid Stars}

\figsetend

\figsetstart
\figsetnum{22}
\figsettitle{Anomalous Cepheid Stars}

\figsetgrpstart
\figsetgrpnum{22}
\figsetgrptitle{Lightcurves for AC-1 to AC-10}
\figsetplot{f22_1.eps}
\figsetgrpnote{Lightcurves of anomalous Cepheids}
\figsetgrpend

\figsetend

\figsetstart
\figsetnum{23}
\figsettitle{Miscellaneous Stars}

\figsetgrpstart
\figsetgrpnum{23.1}
\figsetgrptitle{Lightcurves for Mis-1 to Mis-10}
\figsetplot{f23_1.eps}
\figsetgrpnote{Lightcurves of Miscellaneous}
\figsetgrpend

\figsetend

\clearpage

\begin{deluxetable}{cccrrr}
\tablecolumns{6}
\tablewidth{0pc}
\tablecaption{Coordinates and Number of Repeated Observations in the Carina Fields}
\tablehead{
\colhead{Field} & \colhead{$\alpha$(2000.0)} & \colhead{$\delta$ (2000.0)} &  \colhead{$N_V$} &
\colhead{$N_B$} & \colhead{$N_{\rm nights}$} 
}
\startdata
1 & 06:34:24.2 &  -51:29:15 & 13 & 9   & 2 \\
2 & 06:37:15.3 &  -51:16:35 & 12 & 10 & 3 \\
3 & 06:40:12.7 &  -51:04:39 & 14 & 9   & 3 \\
4 & 06:42:58.8 &  -50:52:24 & 12 & 9   & 3 \\
5 & 06:45:51.4 &  -50:38:28 & 12 & 9   & 2 \\
6 & 06:48:46.7 &  -50:26:37 & 13 & 9   & 3 \\ 
7 & 06:43:00.4 &  -51:04:33 &  2  & 2   & 1 \\
8 & 06:40:14.9 &  -50:52:21 &  2  & 2   & 1 \\
\enddata
\label{tab-fields}
\end{deluxetable}

\begin{deluxetable}{lcccccccccccccc}
\tabletypesize{\scriptsize}
\rotate
\tablecolumns{15}
\tablewidth{0pc}
\tablecaption{Properties of the RR Lyrae Stars}
\tablehead{
ID   &   RA(2000.0) &  DEC(2000.0) &  $\langle B \rangle$ &  $\langle V \rangle$ & Amp B & Amp V &  
$N_B$ & $N_V$ & Period & E(B-V) & Type & ID (D03)\tablenotemark{a} & Type (D03)\tablenotemark{a} & Per (D03)\tablenotemark{a} \\
      &  (deg)            &  (deg)            &                                  &                                 &            &            &
          &             & (d)      &            &          &      &                  & (d) \\
}
\startdata
  RRL-1 &  99.035133 & -51.293030 & 20.94 & 20.55 & 0.55 & 0.45 & 16 & 20 & 0.629 & 0.088 & ab & \nodata & \nodata & \nodata  \\
  RRL-2 &  99.577423 & -51.253571 & 21.04 & 20.60 & 0.83 & 0.65 & 19 & 26 & 0.523 & 0.089 & ab & \nodata & \nodata & \nodata  \\
  RRL-3 &  99.709419 & -51.181721 & 21.06 & 20.62 & 0.56 & 0.46 & 19 & 24 & 0.544 & 0.088 & ab & \nodata & \nodata & \nodata  \\
  RRL-4 &  99.731697 & -51.024300 & 21.32 & 20.85 & 0.17 & 0.24 & 17 & 24 & 0.204 & 0.105 &  c & \nodata & \nodata & \nodata  \\
  RRL-5 &  99.855621 & -51.156158 & 21.27 & 20.82 & 0.14 & 0.17 & 11 & 15 & 0.107 & 0.090 &  c & \nodata & \nodata & \nodata  \\
  RRL-6 & 100.041039 & -51.134750 & 21.21 & 20.78 & 0.35 & 0.29 & 11 & 16 & 0.250 & 0.092 &  c &    V189 &      ab &   0.700  \\
  RRL-7 & 100.122673 & -51.004780 & 21.30 & 20.95 & 0.36 & 0.41 & 11 & 15 & 0.146 & 0.089 &  c &    V151 &       c &   0.343  \\
  RRL-8 & 100.132042 & -50.811119 & 20.49 & 20.35 & 0.72 & 0.56 & 11 & 16 & 0.245 & 0.101 &  c &    V148 &       c &   0.324  \\
  RRL-9 & 100.154999 & -50.990261 & 20.82 & 20.57 & 0.52 & 0.35 & 11 & 16 & 0.246 & 0.088 &  c &    V144 &       c &   0.391  \\
 RRL-10 & 100.222366 & -50.981449 & 20.84 & 20.63 & 0.55 & 0.47 & 11 & 16 & 0.303 & 0.087 &  c &    V125 &      ab &   0.597  \\
 RRL-11 & 100.273499 & -51.007790 & 21.32 & 20.86 & 0.56 & 0.57 & 21 & 27 & 0.679 & 0.083 & ab &    V116 &      ab &   0.685  \\
 RRL-12 & 100.293961 & -51.056358 & 20.92 & 20.67 & 0.68 & 0.51 & 21 & 28 & 0.281 & 0.080 &  c &    V198 &       d &   0.402  \\
 RRL-13 & 100.297867 & -51.037788 & 21.04 & 20.61 & 0.62 & 0.55 & 19 & 27 & 0.663 & 0.081 & ab &    V196 &      ab &   0.670  \\
 RRL-14 & 100.318878 & -51.150841 & 20.94 & 20.62 & 0.84 & 0.65 & 21 & 28 & 0.630 & 0.076 & ab &    V105 &      ab &   0.630  \\
 RRL-15 & 100.336128 & -51.008739 & 21.05 & 20.71 & 0.80 & 0.61 & 21 & 29 & 0.719 & 0.081 & ab &    V192 &       d &   0.390  \\
 RRL-16 & 100.359253 & -50.992741 & 21.14 & 20.73 & 0.47 & 0.42 & 20 & 26 & 0.651 & 0.082 & ab &    V191 &      ab &   0.667  \\
 RRL-17 & 100.364998 & -50.933701 & 20.86 & 20.59 & 0.83 & 0.66 & 22 & 29 & 0.617 & 0.087 & ab &    V183 &      ab &   0.611  \\
 RRL-18 & 100.369789 & -51.114010 & 21.12 & 20.76 & 0.61 & 0.47 & 22 & 29 & 0.628 & 0.073 & ab &     V92 &      ab &   0.620  \\
 RRL-19 & 100.372803 & -51.074558 & 20.76 & 20.44 & 0.43 & 0.32 & 14 & 17 & 0.715 & 0.075 & ab &     V91 &      ab &   0.720  \\
 RRL-20 & 100.374657 & -50.869629 & 21.02 & 20.67 & 0.77 & 0.66 & 20 & 27 & 0.643 & 0.096 & ab &     V90 &      ab &   0.618  \\
 RRL-21 & 100.374786 & -50.787121 & 21.00 & 20.68 & 0.80 & 0.61 & 22 & 29 & 0.383 & 0.104 & ab &     V89 &       d &   0.385  \\
 RRL-22 & 100.398537 & -50.835381 & 21.23 & 20.77 & 0.34 & 0.34 & 22 & 28 & 0.638 & 0.101 & ab &     V85 &      ab &   0.644  \\
 RRL-23 & 100.413544 & -51.094349 & 20.93 & 20.72 & 1.32 & 1.05 & 21 & 30 & 0.602 & 0.073 & ab &     V77 &      ab &   0.605  \\
 RRL-24 & 100.420128 & -51.026020 & 21.23 & 20.80 & 0.66 & 0.56 & 20 & 27 & 0.620 & 0.077 & ab &    V195 &      ab &   0.628  \\
 RRL-25 & 100.428787 & -50.980412 & 20.96 & 20.76 & 1.42 & 1.17 & 22 & 30 & 0.595 & 0.081 & ab & \nodata & \nodata & \nodata  \\
 RRL-26 & 100.444794 & -50.977589 & 21.19 & 20.78 & 0.52 & 0.52 & 22 & 30 & 0.669 & 0.081 & ab &    V200 &      ab &   0.624  \\
 RRL-27 & 100.445541 & -51.118710 & 21.20 & 20.76 & 0.94 & 0.87 & 22 & 30 & 0.566 & 0.072 & ab &     V73 &      ab &   0.571  \\
 RRL-28 & 100.454826 & -50.988541 & 21.12 & 20.71 & 0.57 & 0.48 & 22 & 30 & 0.664 & 0.080 & ab &     V68 &      ab &   0.675  \\
 RRL-29 & 100.455872 & -50.903111 & 21.14 & 20.72 & 0.29 & 0.21 & 22 & 29 & 0.394 & 0.088 &  c &    V179 &      ab &   0.665  \\
 RRL-30 & 100.469383 & -51.088799 & 21.06 & 20.77 & 0.78 & 0.62 & 22 & 30 & 0.601 & 0.073 & ab &     V67 &      ab &   0.613  \\
 RRL-31 & 100.482826 & -50.926441 & 20.94 & 20.56 & 0.99 & 0.80 & 19 & 25 & 0.650 & 0.085 & ab &     V65 &      ab &   0.642  \\
 RRL-32 & 100.492867 & -50.793819 & 21.12 & 20.80 & 0.59 & 0.46 & 20 & 21 & 0.625 & 0.102 & ab &     V61 &      ab &   0.624  \\
 RRL-33 & 100.499001 & -51.110691 & 21.30 & 20.80 & 0.72 & 0.76 & 22 & 30 & 0.603 & 0.072 & ab &     V60 &      ab &   0.615  \\
 RRL-34 & 100.504372 & -50.955441 & 21.01 & 20.71 & 0.92 & 0.81 & 22 & 30 & 0.617 & 0.081 & ab & \nodata & \nodata & \nodata  \\
 RRL-35 & 100.511917 & -50.882431 & 21.00 & 20.71 & 0.93 & 0.77 & 22 & 30 & 0.610 & 0.087 & ab &     V57 &      ab &   0.612  \\
 RRL-36 & 100.517754 & -51.161480 & 21.04 & 20.78 & 1.24 & 1.06 & 21 & 30 & 0.586 & 0.070 & ab &    V206 &      ab &   0.585  \\
 RRL-37 & 100.538002 & -50.898159 & 21.22 & 20.93 & 0.34 & 0.27 & 12 & 14 & 0.246 & 0.085 &  c &     V47 &       c &   0.324  \\
 RRL-38 & 101.874420 & -50.419701 & 21.18 & 20.77 & 0.41 & 0.31 & 16 & 23 & 0.189 & 0.122 &  c & \nodata & \nodata & \nodata  \\
\enddata
\label{tab-RR}
\tablenotetext{a}{Data from \citet{dallora03}.}
\end{deluxetable}

\begin{deluxetable}{lcccccccccccc}
\tabletypesize{\scriptsize}
\rotate
\tablecolumns{13}
\tablewidth{0pc}
\tablecaption{Properties of the anomalous Cepheid Stars}
\tablehead{
ID   &   RA(2000.0) &  DEC(2000.0) &  $\langle B \rangle$ &  $\langle V \rangle$ & Amp B & Amp V &  
$N_B$ & $N_V$ &  Period & E(B-V) & ID (D03)\tablenotemark{a}  & Per (D03)\tablenotemark{a} \\
  &   (deg) &  (deg) &   &  & & &  &  &  (d) & & & (d)  \\
}
\startdata
   AC-1 &    99.185699 &   -51.036240 &   20.44 &   19.95 &   0.21 &   0.20 &   10 &   12 &   0.186 &   0.045 &   \nodata &   \nodata \\ 
   AC-2 &   100.367668 &   -51.012611 &   19.52 &   19.28 &   0.25 &   0.24 &   22 &   28 &   0.444 &   0.061 &      V193 &     0.424 \\ 
   AC-3 &   100.386169 &   -50.950211 &   19.00 &   18.73 &   0.43 &   0.40 &   19 &   24 &   0.859 &   0.060 &       V87 &     0.880 \\ 
   AC-4 &   100.415833 &   -50.983921 &   19.53 &   19.20 &   1.38 &   1.19 &   22 &   30 &   0.545 &   0.063 &      V190 &     1.160 \\ 
   AC-5 &   100.433540 &   -50.838100 &   19.29 &   18.99 &   0.95 &   0.77 &   22 &   30 &   0.502 &   0.059 &      V178 &     0.507 \\ 
   AC-6 &   100.483871 &   -50.989330 &   19.25 &   19.09 &   0.51 &   0.72 &   20 &   29 &   0.312 &   0.066 &      V203 &     0.467 \\ 
   AC-7 &   100.602158 &   -51.044540 &   19.89 &   19.68 &   0.19 &   0.16 &   11 &   14 &   0.197 &   0.070 &      V205 &     0.383 \\ 
   AC-8 &   100.642174 &   -50.972210 &   19.04 &   18.78 &   0.55 &   0.39 &   10 &   12 &   0.161 &   0.066 &   \nodata &   \nodata \\ 
   AC-9 &   101.818748 &   -50.474560 &   19.21 &   19.13 &   0.36 &   0.55 &    8 &   23 &   0.476 &   0.091 &   \nodata &   \nodata \\ 
  AC-10 &   102.226418 &   -50.509750 &   19.71 &   19.35 &   0.18 &   0.32 &    9 &   13 &   0.163 &   0.089 &   \nodata &   \nodata \\ 
\enddata
\label{tab-AC}
\tablenotetext{a}{Data from \citet{dallora03}.}
\end{deluxetable}

\begin{deluxetable}{lcccccccccccccc}
\tabletypesize{\scriptsize}
\rotate
\tablecolumns{15}
\tablewidth{0pc}
\tablecaption{Properties of the Dwarf Cepheid Stars}
\tablehead{
ID   &   RA(2000.0) &  DEC(2000.0) &  $\langle B \rangle$ &  $\langle V \rangle$ & Amp B & Amp V &  
$N_B$ & $N_V$ &  Period & Alt Period & E(B-V) & Mode & ID (M98)\tablenotemark{a}  & Per (M98)\tablenotemark{a} \\
  &   (deg) &  (deg) &   &  & & &  &  &  (d) & (d) &  & & & (d)  \\
}
\startdata
  DC-44  & 100.226624 & -51.024540 & 22.27 & 22.05 & 0.50 & 0.57 & 11 & 14 & 0.07410 &  \nodata & 0.054 & \nodata & \nodata & \nodata \\
  DC-45  & 100.231827 & -51.011921 & 23.41 & 23.06 & 0.56 & 0.60 &  8 & 15 & 0.05779 & 0.05950 & 0.054   &     F  &     V4 & 0.05991 \\
  DC-46 & 100.240211 & -50.935970 & 23.02 & 22.76 & 0.42 & 0.38 &  9 & 16 & 0.06229 & \nodata  & 0.054 & \nodata & \nodata & \nodata \\
  DC-47 & 100.241219 & -51.035511 & 23.23 & 22.91 & 0.64 & 0.68 &  9 & 15 & 0.05629 & \nodata & 0.055 &       F & \nodata  & \nodata \\
  DC-48 & 100.246582 & -50.828979 & 23.24 & 23.07 & 0.59 & 0.42 & 11 & 16 & 0.05489 & \nodata & 0.053  &       F & \nodata & \nodata \\
\enddata
\label{tab-DC}
\tablenotetext{a}{Data from \citet{mateo98}.}
\tablecomments{Table~\ref{tab-DC} is published in its entirety in the electronic edition of the journal. 
A portion is shown here for guidance regarding its form and content.}
\end{deluxetable}

\begin{deluxetable}{lccccccccccc}
\tabletypesize{\scriptsize}
\rotate
\tablecolumns{12}
\tablewidth{0pc}
\tablecaption{Properties of the Miscellaneous Periodic Variable Stars}
\tablehead{
ID   &   RA(2000.0) &  DEC(2000.0) &  $\langle B \rangle$ &  $\langle V \rangle$ & Amp B & Amp V &  
$N_B$ & $N_V$ &  Period & E(B-V) & Comment\\
  &   (deg) &  (deg) &   &  & & &  &  &  (d) &  & \\
}
\startdata
  Mis-1 &    99.709000 &   -51.342861 &   21.56 &   21.13 &   0.19 &   0.16 &   19 &   24 &   0.144 &   0.047 & \\ 
  Mis-2 &   100.265793 &   -50.914619 &   22.40 &   22.36 &   0.37 &   0.34 &   18 &   24 &   0.234 &   0.054 & \\ 
  Mis-3 &   100.276123 &   -50.924320 &   22.94 &   22.78 &   0.46 &   0.47 &   18 &   26 &   0.245 &   0.055 & \\ 
  Mis-4 &   100.291924 &   -50.987759 &   23.71 &   23.27 &   0.60 &   0.63 &   16 &   22 &   0.227 &   0.056 & \\ 
  Mis-5 &   100.322502 &   -51.095890 &   22.01 &   21.82 &   0.38 &   0.35 &   21 &   29 &   0.204 &   0.061 & Eclipsing Binary \\ 
  Mis-6 &   100.327461 &   -50.826900 &   23.71 &   23.49 &   0.57 &   0.49 &   20 &   24 &   0.216 &   0.054 & \\ 
  Mis-7 &   100.421043 &   -50.957130 &   23.25 &   23.16 &   0.41 &   0.40 &   17 &   25 &   0.173 &   0.062 & \\ 
  Mis-8 &   101.777763 &   -50.612900 &   23.63 &   23.49 &   0.88 &   0.87 &   12 &   18 &   0.238 &   0.086 & \\ 
  Mis-9 &   101.782082 &   -50.743198 &   23.94 &   23.36 &   0.72 &   0.54 &    7 &   17 &   0.193 &   0.098 & \\ 
\enddata
\label{tab-Mis}
\end{deluxetable}

\begin{deluxetable}{lccrrr}
\tablecolumns{4}
\tablewidth{0pc}
\tablecaption{Dereddened Distance Modulus for Carina using different P-L relationships for DC stars}
\tablehead{
\colhead{P-L relationship} & \colhead{[Fe/H]} & \colhead{$\mu_0$} & \colhead{rms} }
\startdata
\citet{nemec94} 		  &	-2.0 	&	19.99	&	0.10	\\
								  &	-1.7	& 	19.89	& 	0.10	\\
\citet{poretti08} 		  &		     & 	20.20	& 0.13  \\
\citet{mcnamara11} 	  &	-2.0  &	20.18	& 0.10  \\
								  &	-1.7	&	20.24	& 0.10	\\
\citet{cohen12} 		      &		         & 20.33    &  0.12  \\
\tableline
Mean (for [Fe/H]=-1.7) &             & 20.17    & 0.10 \\
\enddata
\label{tab-distance}
\end{deluxetable}

\begin{deluxetable}{lrrr}
\tablecolumns{4}
\tablewidth{0pc}
\tablecaption{Specific Frequency of DC in Local Group objects and globular clusters}
\tablehead{
\colhead{Object} & \colhead{$M_V$\tablenotemark{a}} & \colhead{$N_{\rm DC}$\tablenotemark{b}} &  \colhead{$S_{\rm DC}$}
}
\startdata
Carina     & -9.3   & 315   & 73  \\
Fornax    & -13.2 & 97     & 0.6 \\
LMC        & -18.5 & 2323 & 0.1\tablenotemark{c}  \\
\tableline
NGC 288 & -6.75 & 6       & 15 \\
NGC4372 & -7.79	&  8 	&	10 \\
NGC5053	& -6.76	 & 5 	&	13 \\
NGC5139 ($\omega$Cen)	 & -10.26	&  34 	&	3	\\
NGC5466	&  -6.98 & 9 	&	19	 \\
NGC5904 (M5) & -8.81 &	5 &	2	\\
NGC6809 (M55) &  -7.57 &	27 	&	33 \\	
\enddata
\label{tab-frequency}
\tablenotetext{a}{Values for Carina and Fornax from \citet{mateo98rev}, LMC from 
\citet{binney98}, and globular clusters from the online (2010) version of \citet{harris96}
catalog.}
\tablenotetext{b}{The values of $N_{\rm DC}$ 
for Carina and Fornax have been corrected to include all
stars within the tidal King radius (see text). No correction was applied to the LMC or the 
globular clusters}
\tablenotetext{c}{Uncertain}
\end{deluxetable}


\begin{thebibliography}

\bibitem[Balona \& Dziembowski(2011)]{balona11} Balona, L.~A., \& Dziembowski, W.~A.\ 2011, \mnras, 417, 591 

\bibitem[Balona \& Nemec(2012)]{balona12} Balona, L.~A., \& Nemec, J.~M.\ 2012, \mnras, 426, 2413 

\bibitem[{{Battaglia} {et~al.}(2012){Battaglia}, {Irwin}, {Tolstoy}, {de Boer},
  \& {Mateo}}]{battaglia12}
{Battaglia}, G., {Irwin}, M., {Tolstoy}, E., {de Boer}, T., \& {Mateo}, M.
  2012, \apjl, 761, L31

\bibitem[{{Binney} \& {Merrifield}(1998)}]{binney98}
{Binney}, J., \& {Merrifield}, M. 1998, {Galactic Astronomy}

\bibitem[{{Bono} {et~al.}(1997){Bono}, {Caputo}, {Santolamazza}, {Cassisi}, \&
  {Piersimoni}}]{bono97}
{Bono}, G., {Caputo}, F., {Santolamazza}, P., {Cassisi}, S., \& {Piersimoni},
  A. 1997, \aj, 113, 2209

\bibitem[{{Bono} {et~al.}(2010){Bono}, {Stetson}, {Walker}, {Monelli},
  {Fabrizio}, {Pietrinferni}, {Brocato}, {Buonanno}, {Caputo}, {Cassisi},
  {Castellani}, {Cignoni}, {Corsi}, {Dall'Ora}, {Degl'Innocenti}, {Fran{\c
  c}ois}, {Ferraro}, {Iannicola}, {Nonino}, {Moroni}, {Pulone}, {Smith}, \&
  {Thevenin}}]{bono10}
{Bono}, G. {et~al.} 2010, \pasp, 122, 651

\bibitem[{{Breger}(2000)}]{breger00}
{Breger}, M. 2000, in Astronomical Society of the Pacific Conference Series,
  Vol. 210, Delta Scuti and Related Stars, ed. M.~{Breger} \& M.~{Montgomery},
  3

\bibitem[{{Clement} {et~al.}(2001){Clement}, {Muzzin}, {Dufton}, {Ponnampalam},
  {Wang}, {Burford}, {Richardson}, {Rosebery}, {Rowe}, \& {Hogg}}]{clement01}
{Clement}, C.~M. {et~al.} 2001, \aj, 122, 2587

\bibitem[{{Cohen} \& {Sarajedini}(2012)}]{cohen12}
{Cohen}, R.~E., \& {Sarajedini}, A. 2012, \mnras, 419, 342

\bibitem[{{Coleman} \& {de Jong}(2008)}]{coleman08}
{Coleman}, M.~G., \& {de Jong}, J.~T.~A. 2008, \apj, 685, 933

\bibitem[{{Dall'Ora} {et~al.}(2003){Dall'Ora}, {Ripepi}, {Caputo},
  {Castellani}, {Bono}, {Smith}, {Brocato}, {Buonanno}, {Castellani}, {Corsi},
  {Marconi}, {Monelli}, {Nonino}, {Pulone}, \& {Walker}}]{dallora03}
{Dall'Ora}, M. {et~al.} 2003, \aj, 126, 197

\bibitem[{{Fiorentino} {et~al.}(2012){Fiorentino}, {Stetson}, {Monelli},
  {Bono}, {Bernard}, \& {Pietrinferni}}]{fiorentino12}
{Fiorentino}, G., {Stetson}, P.~B., {Monelli}, M., {Bono}, G., {Bernard},
  E.~J., \& {Pietrinferni}, A. 2012, \apjl, 759, L12

\bibitem[{{Garg} {et~al.}(2010){Garg}, {Cook}, {Nikolaev}, {Huber}, {Rest},
  {Becker}, {Challis}, {Clocchiatti}, {Miknaitis}, {Minniti}, {Morelli},
  {Olsen}, {Prieto}, {Suntzeff}, {Welch}, \& {Wood-Vasey}}]{garg10}
{Garg}, A. {et~al.} 2010, \aj, 140, 328

\bibitem[{{Grebel}(1999)}]{grebel99}
{Grebel}, E.~K. 1999, in IAU Symposium, Vol. 192, The Stellar Content of Local
  Group Galaxies, ed. P.~{Whitelock} \& R.~{Cannon}, 17

\bibitem[{{Grebel}(2011)}]{grebel11}
{Grebel}, E.~K. 2011, in EAS Publications Series, Vol.~48, EAS Publications
  Series, ed. M.~{Koleva}, P.~{Prugniel}, \& I.~{Vauglin}, 315--327

\bibitem[{{Harris}(1996)}]{harris96}
{Harris}, W.~E. 1996, \aj, 112, 1487

\bibitem[{{Helmi} {et~al.}(2006){Helmi}, {Irwin}, {Tolstoy}, {Battaglia},
  {Hill}, {Jablonka}, {Venn}, {Shetrone}, {Letarte}, {Arimoto}, {Abel},
  {Francois}, {Kaufer}, {Primas}, {Sadakane}, \& {Szeifert}}]{helmi06}
{Helmi}, A. {et~al.} 2006, \apjl, 651, L121

\bibitem[{{Hurley-Keller} {et~al.}(1998){Hurley-Keller}, {Mateo}, \&
  {Nemec}}]{hurley98}
{Hurley-Keller}, D., {Mateo}, M., \& {Nemec}, J. 1998, \aj, 115, 1840

\bibitem[{{Irwin} \& {Hatzidimitriou}(1995)}]{irwin95}
{Irwin}, M., \& {Hatzidimitriou}, D. 1995, \mnras, 277, 1354

\bibitem[{{Jeon} {et~al.}(2004){Jeon}, {Lee}, {Kim}, \& {Lee}}]{jeon04}
{Jeon}, Y.-B., {Lee}, M.~G., {Kim}, S.-L., \& {Lee}, H. 2004, \aj, 128, 287

\bibitem[{{Kinemuchi} {et~al.}(2008){Kinemuchi}, {Harris}, {Smith},
  {Silbermann}, {Snyder}, {La Cluyz{\'e}}, \& {Clark}}]{kinemuchi08}
{Kinemuchi}, K., {Harris}, H.~C., {Smith}, H.~A., {Silbermann}, N.~A.,
  {Snyder}, L.~A., {La Cluyz{\'e}}, A.~P., \& {Clark}, C.~L. 2008, \aj, 136,
  1921

\bibitem[{{King}(1962)}]{king62}
{King}, I. 1962, \aj, 67, 471

\bibitem[{{Koch} {et~al.}(2006){Koch}, {Grebel}, {Wyse}, {Kleyna}, {Wilkinson},
  {Harbeck}, {Gilmore}, \& {Evans}}]{koch06}
{Koch}, A., {Grebel}, E.~K., {Wyse}, R.~F.~G., {Kleyna}, J.~T., {Wilkinson},
  M.~I., {Harbeck}, D.~R., {Gilmore}, G.~F., \& {Evans}, N.~W. 2006, \aj, 131,
  895

\bibitem[{{Kuehn} {et~al.}(2008){Kuehn}, {Kinemuchi}, {Ripepi}, {Clementini},
  {Dall'Ora}, {Di Fabrizio}, {Rodgers}, {Greco}, {Marconi}, {Musella}, {Smith},
  {Catelan}, {Beers}, \& {Pritzl}}]{kuehn08}
{Kuehn}, C. {et~al.} 2008, \apjl, 674, L81

\bibitem[{{Lafler} \& {Kinman}(1965)}]{lafler65}
{Lafler}, J., \& {Kinman}, T.~D. 1965, \apjs, 11, 216

\bibitem[{{Majewski} {et~al.}(2000){Majewski}, {Ostheimer}, {Patterson},
  {Kunkel}, {Johnston}, \& {Geisler}}]{majewski00}
{Majewski}, S.~R., {Ostheimer}, J.~C., {Patterson}, R.~J., {Kunkel}, W.~E.,
  {Johnston}, K.~V., \& {Geisler}, D. 2000, \aj, 119, 760

\bibitem[{{Mateo}(1993)}]{mateo93}
{Mateo}, M. 1993, in Astronomical Society of the Pacific Conference Series,
  Vol.~53, Blue Stragglers, ed. R.~A. {Saffer}, 74

\bibitem[{{Mateo} {et~al.}(1995){Mateo}, {Fischer}, \& {Krzeminski}}]{mateo95}
{Mateo}, M., {Fischer}, P., \& {Krzeminski}, W. 1995, \aj, 110, 2166

\bibitem[{{Mateo} {et~al.}(1998){Mateo}, {Hurley-Keller}, \& {Nemec}}]{mateo98}
{Mateo}, M., {Hurley-Keller}, D., \& {Nemec}, J. 1998, \aj, 115, 1856

\bibitem[{{Mateo}(1998)}]{mateo98rev}
{Mateo}, M.~L. 1998, \araa, 36, 435

\bibitem[{{Mateu} {et~al.}(2012){Mateu}, {Vivas}, {Downes}, {Brice{\~n}o},
  {Zinn}, \& {Cruz-Diaz}}]{mateu12}
{Mateu}, C., {Vivas}, A.~K., {Downes}, J.~J., {Brice{\~n}o}, C., {Zinn}, R., \&
  {Cruz-Diaz}, G. 2012, \mnras, 427, 3374

\bibitem[{{McNamara}(2011)}]{mcnamara11}
{McNamara}, D.~H. 2011, \aj, 142, 110

\bibitem[{{Monelli} {et~al.}(2003){Monelli}, {Pulone}, {Corsi}, {Castellani},
  {Bono}, {Walker}, {Brocato}, {Buonanno}, {Caputo}, {Castellani}, {Dall'Ora},
  {Marconi}, {Nonino}, {Ripepi}, \& {Smith}}]{monelli03}
{Monelli}, M. {et~al.} 2003, \aj, 126, 218

\bibitem[{{Mu{\~n}oz} {et~al.}(2006){Mu{\~n}oz}, {Majewski}, {Zaggia},
  {Kunkel}, {Frinchaboy}, {Nidever}, {Crnojevic}, {Patterson}, {Crane},
  {Johnston}, {Sohn}, {Bernstein}, \& {Shectman}}]{munoz06}
{Mu{\~n}oz}, R.~R. {et~al.} 2006, \apj, 649, 201

\bibitem[{{Nemec} {et~al.}(1994){Nemec}, {Nemec}, \& {Lutz}}]{nemec94}
{Nemec}, J.~M., {Nemec}, A.~F.~L., \& {Lutz}, T.~E. 1994, \aj, 108, 222

\bibitem[{{Pont} {et~al.}(2004){Pont}, {Zinn}, {Gallart}, {Hardy}, \&
  {Winnick}}]{pont04}
{Pont}, F., {Zinn}, R., {Gallart}, C., {Hardy}, E., \& {Winnick}, R. 2004, \aj,
  127, 840

\bibitem[{{Poretti} {et~al.}(2008){Poretti}, {Clementini}, {Held}, {Greco},
  {Mateo}, {Dell'Arciprete}, {Rizzi}, {Gullieuszik}, \& {Maio}}]{poretti08}
{Poretti}, E. {et~al.} 2008, \apj, 685, 947

\bibitem[{{Pych} {et~al.}(2001){Pych}, {Kaluzny}, {Krzeminski},
  {Schwarzenberg-Czerny}, \& {Thompson}}]{pych01}
{Pych}, W., {Kaluzny}, J., {Krzeminski}, W., {Schwarzenberg-Czerny}, A., \&
  {Thompson}, I.~B. 2001, \aap, 367, 148

\bibitem[{{Rest} {et~al.}(2005){Rest}, {Stubbs}, {Becker}, {Miknaitis},
  {Miceli}, {Covarrubias}, {Hawley}, {Smith}, {Suntzeff}, {Olsen}, {Prieto},
  {Hiriart}, {Welch}, {Cook}, {Nikolaev}, {Huber}, {Prochtor}, {Clocchiatti},
  {Minniti}, {Garg}, {Challis}, {Keller}, \& {Schmidt}}]{rest05}
{Rest}, A. {et~al.} 2005, \apj, 634, 1103

\bibitem[{{Rizzi} {et~al.}(2003){Rizzi}, {Held}, {Bertelli}, \&
  {Saviane}}]{rizzi03}
{Rizzi}, L., {Held}, E.~V., {Bertelli}, G., \& {Saviane}, I. 2003, \apjl, 589,
  L85

\bibitem[{{Rizzi} {et~al.}(2007){Rizzi}, {Held}, {Saviane}, {Tully}, \&
  {Gullieuszik}}]{rizzi07}
{Rizzi}, L., {Held}, E.~V., {Saviane}, I., {Tully}, R.~B., \& {Gullieuszik}, M.
  2007, \mnras, 380, 1255

\bibitem[{{Saha} {et~al.}(1986){Saha}, {Monet}, \& {Seitzer}}]{saha86}
{Saha}, A., {Monet}, D.~G., \& {Seitzer}, P. 1986, \aj, 92, 302

\bibitem[{{Schechter} {et~al.}(1993){Schechter}, {Mateo}, \&
  {Saha}}]{schechter93}
{Schechter}, P.~L., {Mateo}, M., \& {Saha}, A. 1993, \pasp, 105, 1342

\bibitem[{{Schlegel} {et~al.}(1998){Schlegel}, {Finkbeiner}, \&
  {Davis}}]{schlegel98}
{Schlegel}, D.~J., {Finkbeiner}, D.~P., \& {Davis}, M. 1998, \apj, 500, 525

\bibitem[{{Smecker-Hane} {et~al.}(1996){Smecker-Hane}, {Stetson}, {Hesser}, \&
  {Vandenberg}}]{smecker96}
{Smecker-Hane}, T.~A., {Stetson}, P.~B., {Hesser}, J.~E., \& {Vandenberg},
  D.~A. 1996, in Astronomical Society of the Pacific Conference Series,
  Vol.~98, From Stars to Galaxies: the Impact of Stellar Physics on Galaxy
  Evolution, ed. C.~{Leitherer}, U.~{Fritze-von-Alvensleben}, \& J.~{Huchra},
  328
  
\bibitem[{{Smith}(1995)}]{smith95}
{Smith}, H.~A. 1995, {RR Lyrae stars} (Cambridge Astrophysics Series,
  Cambridge, New York: Cambridge University Press, |c1995)

\bibitem[{{Sollima} {et~al.}(2010){Sollima}, {Cacciari}, {Bellazzini}, \&
  {Colucci}}]{sollima10}
{Sollima}, A., {Cacciari}, C., {Bellazzini}, M., \& {Colucci}, S. 2010, \mnras,
  406, 329

\bibitem[{{Stetson}(1996)}]{stetson96}
{Stetson}, P.~B. 1996, \pasp, 108, 851

\bibitem[{{Vivas} \& {Zinn}(2006)}]{vivas06}
{Vivas}, A.~K., \& {Zinn}, R. 2006, \aj, 132, 714

\bibitem[{{Vivas} {et~al.}(2004){Vivas}, {Zinn}, {Abad}, {Andrews}, {Bailyn},
  {Baltay}, {Bongiovanni}, {Brice{\~n}o}, {Bruzual}, {Coppi}, {Della Prugna},
  {Ellman}, {Ferr{\'{\i}}n}, {Gebhard}, {Girard}, {Hernandez}, {Herrera},
  {Honeycutt}, {Magris}, {Mufson}, {Musser}, {Naranjo}, {Rabinowitz},
  {Rengstorf}, {Rosenzweig}, {S{\'a}nchez}, {S{\'a}nchez}, {Schaefer},
  {Schenner}, {Snyder}, {Sofia}, {Stock}, {van Altena}, {Vicente}, \&
  {Vieira}}]{vivas04}
{Vivas}, A.~K. {et~al.} 2004, \aj, 127, 1158

\bibitem[{{Walcher} {et~al.}(2003){Walcher}, {Fried}, {Burkert}, \&
  {Klessen}}]{walcher03}
{Walcher}, C.~J., {Fried}, J.~W., {Burkert}, A., \& {Klessen}, R.~S. 2003,
  \aap, 406, 847

\bibitem[{{Watkins} {et~al.}(2009){Watkins}, {Evans}, {Belokurov}, {Smith},
  {Hewett}, {Bramich}, {Gilmore}, {Irwin}, {Vidrih}, {Wyrzykowski}, \&
  {Zucker}}]{watkins09}
{Watkins}, L.~L. {et~al.} 2009, \mnras, 398, 1757

\bibitem[{{Zinn} \& {Searle}(1976)}]{zinn76}
{Zinn}, R., \& {Searle}, L. 1976, \apj, 209, 734

\end{thebibliography}
\end{document}